\documentclass[a4paper,10pt]{article}

\usepackage{bm}
\usepackage{verbatim}
\usepackage[colorlinks=false]{hyperref} % citecolor=blue ?l????????
\usepackage{subfigure}
\usepackage{graphics}
\usepackage{amsmath,amsfonts,amssymb,graphics,graphicx,epsfig,color,times}
\usepackage{pgfplots}
\usepackage{verbatim}
\usepackage{epstopdf}
\usepackage{mathrsfs}
\usepackage{multirow}

\def\picill#1by#2(#3)
{\vbox to #2 {\hrule width #1 height 0pt depth 0pt
\vfill\epsffile{#3}}}

\newtheorem{theorem}{\hspace{2em}Theorem}[section]

\newtheorem{lemma}{\hspace{2em}Lemma}[section]
\newtheorem{corollary}{\hspace{2em}Corollary}[section]

\begin{document}
\setlength{\unitlength}{1mm}

\thispagestyle{empty}

 \begin{center}
%{Notes on Bell states and quantum teleportation}
Algebraic and Topological Study of Bell States and Quantum Teleportation
  \\[4mm]

\vspace{.5cm}
Yong Zhang ${}^{a,}$\footnote{yong\_zhang@whu.edu.cn},
Wei Zeng ${}^{a,}$\footnote{zengwei@whu.edu.cn}, 
Ming Lian ${}^{a,}$\footnote{m.lien@whu.edu.cn}
\\[.5cm]

${}^a$ School of Physics and Technology, Wuhan University, P.R. China 430072
\\[0.1cm]
\end{center}

\vspace{0.2cm}

\begin{center}
\parbox{13.6cm}{
\centerline{\small  \bf Abstract}  \noindent

Bell states and quantum teleportation play crucial roles in quantum information and computation. However, a comprehensive theoretical study of both topics remains to be carried out.
This work aims to investigate key algebraic properties of generalized Bell states and explore the topological features of quantum teleportation. First, the basis theorem and the basis group are introduced to show that the extension of a generalized Bell basis by a unitary matrix still forms an orthonormal basis. Then, a twist operator is defined to establish a connection between a generalized multi‑qubit Bell state and a tensor product of two‑qubit Bell states. In addition, the Temperley–Lieb algebra, the braid group relations, and the Yang–Baxter equation are employed to provide a topological description of generalized Bell states and quantum teleportation. The results demonstrate that our approach not only offers a clear illustration of relevant quantum information protocols but also reveals the topological nature of quantum entanglement and teleportation.

}

\end{center}

\vspace{.2cm}

\begin{tabbing}

\end{tabbing}

%\newpage
%\pdfbookmark[0]{Contents}{toc}

%\tableofcontents

%\newpage

%%%%%%%%%%%%%%%%%%%
%%%%%%%%%%%%%%%%%%%
%%%%%%%%%%%%%%%%%%%

\newpage

\section{Introduction}

The Bell states, also known as EPR pairs \cite{EPR35, Bohm51, Bell64}, are associated with the great pioneers of quantum mechanics—Einstein, Bohm, and Bell. In particular, Bell's inequalities deny the existence of local deterministic theories and justify nonlocal quantum theory. As the prototypical maximally entangled pure states in quantum entanglement theory \cite{HHHH2009, FVMH2019, EKZ2020, MLS2025, Hill1997, Coffman2000, Wong2001}, they provide a crucial resource for performing quantum teleportation \cite{Bennett_1993, BPMFWZ1997, BBDHP1998, Ren2017}, which is regarded as one of the most important quantum information protocols. By combining Bell states, Bell measurements, classical communication, and unitary corrections, quantum teleportation enables an unknown qubit to be sent from Alice to Bob, who are spatially separated and do not share a quantum communication channel. It should be noted that Alice and Bob were initially co-located at the beginning of the experiment to create the Bell states, and that a classical channel is subsequently required to carry out the correction. The peculiar properties of Bell states and quantum teleportation not only raise the fundamental question of why they work so successfully in nature—that is, why quantum mechanics has proved so remarkably accurate—but also serve as powerful tools for investigating a wide range of topics in quantum information and computation.

Bell states admit a Schmidt decomposition in a two‑qubit state space \cite{NC2011}, and thus can be directly generalized in several ways. Generalized two‑qudit Bell states are defined in a two‑qudit Hilbert space, while generalized multiple qubit Bell states are defined in a multiple qubit Hilbert space; collectively, these are referred to as generalized Bell states. An orthonormal basis consisting of generalized Bell states is called a generalized Bell basis. Quantum teleportation via generalized Bell states has been extensively studied in the literature, and a rich body of work exists on generalized Bell states and quantum teleportation. The standard textbook \cite{NC2011} and two review articles \cite{PEWFB2015, HGLLG2023} serve as excellent references for a comprehensive introduction to the properties and applications of Bell states and quantum teleportation. It is worth noting that references on quantum teleportation are often also the best references for Bell states, and thus it is natural and appropriate to study both topics in the same paper.

To highlight the relevance and interest of our study of Bell states and quantum teleportation, we briefly survey a broad range of applications stemming from the original teleportation proposal \cite{Bennett_1993}. First, the connection between Bell states and coherent states has been investigated in \cite{Fujii2001, Fivel2002}, while a generalized Bell basis has been explored in \cite{SL2009}. Subsequently, teleportation of a single qudit via generalized two‑qudit Bell states has been addressed in \cite{BDM00, Fujii2001a, Werner01, Gay_2014, Fonseca_2019, Zhang_2019, Luo_2019, Hu_2020, DM2025}; teleportation of multiple qubits using generalized multiple qubit Bell states has been systematically presented in \cite{Lee_2002, Rigolin_2005, Chen_2006, Yeo_2006, CZ2009, ZLLF2012}; teleportation with non‑maximally entangled states has been discussed in \cite{MH1999, Bandyopadhyay2000, Son_2001, RDF2003, KCL_2004}; and teleportation with continuous variables was first addressed in \cite{Vaidman94, BK1998, FSBFKP1998}.

Furthermore, quantum circuits for teleportation \cite{BBC1998} have been applied in the study of teleportation based quantum computation \cite{GC1999, KLM2001, Nielsen2003, Leung2004, HRZDG2004, Gao2010, EL2019, Chou2018, Wan2019, Daiss2021} and one way quantum computation \cite{RB2001, Nielsen2004, GE2007, BBDR2009, WRC2024}, both of which are universal models of quantum computation and can be unified within the framework of measurement based quantum computation \cite{AL2004, CLN2004, Jozsa2006}. In addition to applications in quantum computation, the combination of a variant of quantum teleportation, known as entanglement swapping \cite{ZZHE1993}, with entanglement distillation \cite{Bennett1996} enables long distance quantum communication through so called quantum repeaters \cite{BCZ1998}. Moreover, quantum teleportation with quantum memories is being used to create quantum networks or a quantum internet \cite{Kimble2008, WEH2018, DKD2018, Langenfeld2021, Hermans2022}. Quantum teleportation via multipartite entangled states is also referred to as quantum teleportation networking, including controlled quantum teleportation \cite{KB1998, PP2019, BCL2019}, quantum secret sharing \cite{HBB1999, LLJP2020}, and tele‑cloning \cite{Bruss1998}.

Additionally, the literature contains many other important applications of quantum teleportation, including port based quantum teleportation \cite{IH2008}, entanglement estimation \cite{SSC2019}, teleportation with spin chains \cite{Apollaro2019}, teleportation as an inverse of measurement \cite{LIKNK2021}, teleportation with one classical bit \cite{Parakh2022}, distributed quantum computing \cite{Main2025}, teleportation with multiple degrees of freedom \cite{Wang2015, Ru2021}, teleportation with identical particles \cite{Marzolino2015, Debarba2020}, and nonclassical teleportation \cite{CSS2017, Carvacho2018, CCHL2021, Hu2019, LS2020}, among others. In particular, the algebraic and topological properties underlying quantum teleportation, as discussed in \cite{AC04, Kauffman2005, Zhang2006, Zhang2009, ZZP2015, ZZ2017}, constitute the central focus of the present paper.

As noted above, Bell states and quantum teleportation play an important role in quantum mechanics and quantum information science. Nevertheless, a comprehensive and progressive investigation of both remains a highly valuable research endeavor. This paper highlights crucial algebraic properties of generalized Bell bases and provides a topological study of generalized Bell states and quantum teleportation. Two typical examples of generalized Bell bases are a generalized two‑qudit Bell basis constructed from products of generalized Pauli matrices \cite{DM2025, RDF2003, PZ1988} and a generalized multiple qubit Bell basis formed from tensor products of Pauli matrices. In addition, the extension of a generalized Bell basis by a local unitary matrix is analyzed in detail.

The paper presents several innovative approaches to generalized Bell states. First, the basis theorem, which is primarily motivated by teleportation based quantum computation \cite{GC1999} and one way quantum computation \cite{RB2001}, can be viewed as a generalization of Kauffman's basis lemma \cite{Kauffman2005}. We propose and verify that the extension of a generalized Bell basis by the local action of a matrix remains an orthonormal basis if and only if the matrix is unitary. In particular, a novel concept—the basis group—is introduced to achieve this result. Second, we define a twist operator and explore its algebraic properties, including its decomposition as a product of permutations, in order to establish a clear correspondence between generalized two‑qudit Bell states and generalized multiple qubit Bell states. Third, we construct a complete set of observables that have generalized two‑qudit Bell states as their eigenvectors. We also investigate a quantum circuit model, a complete set of observables, and quantum entanglement for generalized multiple qubit Bell states. In addition, we formulate a generalized concurrence \cite{Rigolin_2005} for a \(2n\)-qubit pure state.

Kauffman presented a topological description of quantum teleportation in \cite{Kauffman2005}. As discussed in Kauffman's seminal book \cite{Kauffman2012}, the Temperley–Lieb algebra \cite{TL1971}, the braid group, and the Yang–Baxter equation \cite{Yang1967} are three essential tools for the study of low‑dimensional topology. Building on this foundation, the first author of the present paper proposed both the extended Temperley–Lieb diagrammatic approach and the braid teleportation approach \cite{Zhang2006, Zhang2009} to characterize Bell states and quantum teleportation from a topological perspective. In contrast to the standard description of quantum information protocols using generalized Bell states, the extended Temperley–Lieb diagrammatic approach not only provides a concise illustration that captures the essential features of complicated algebraic calculations but also allows us to explore the topological aspects of quantum information and computation. Furthermore, going beyond the traditional understanding of quantum teleportation, the braid teleportation approach treats quantum teleportation in the space of braiding configurations, where each braid corresponds to a Yang–Baxter gate \cite{KL2004, ZKG04, ZKG05}—that is, a two‑qubit quantum gate satisfying the Yang–Baxter equation.

In \cite{Zhang2006, Zhang2009}, the original extended Temperley–Lieb diagrammatic approach was applied to quantum information protocols using generalized two‑qudit Bell states, while the braid teleportation approach was employed to characterize single‑qubit teleportation. However, the present paper advances both approaches to provide a topological interpretation of multi‑qubit quantum teleportation. We show that extending these two approaches by incorporating SWAP gates accomplishes this task. Note that the Temperley–Lieb algebra augmented with SWAP gates yields the Brauer algebra \cite{Brauer1937}, while Yang–Baxter gates combined with SWAP gates generate high‑dimensional Yang–Baxter gates. Our study thus establishes a fundamental connection between quantum information science and low‑dimensional topology, which will motivate further progress in both fields.

The remainder of this paper is organized as follows. Section 2 reviews Bell states and quantum teleportation in the context of quantum information and computation \cite{NC2011}. Section 3 describes generalized two‑qudit Bell states and explains the basis theorem and the basis group. Section 4 presents generalized two‑qudit Bell states under the local action of generalized Pauli matrices and constructs a complete set of observables for which they are eigenvectors. Section 5 provides a systematic study of generalized multiple qubit Bell states, covering topics such as the twist operator, the basis theorem, quantum circuit models, a complete set of observables, and quantum entanglement measures. Section 6 applies the extended Temperley–Lieb diagrammatic approach to generalized two‑qudit Bell states and generalized multiple qubit Bell states, respectively. Section 7 derives various teleportation equations using generalized Bell states and describes quantum teleportation of a single qudit or multiple qubits. Section 8 introduces Yang–Baxter gates, the Bell transform, and braid teleportation, and then combines Yang–Baxter gates with SWAP gates to illustrate braid teleportation of multiple qubits. Section 9 concludes the paper with remarks on future research. Appendix A contains detailed calculations involving tensor products of Pauli matrices.

\section{Notation for Bell states and quantum teleportation}

The concepts, notations, and conventions of quantum information and computation are adopted from the standard textbook \cite{NC2011}.

A qubit is described by a two-dimensional Hilbert space $\mathscr{H}_2$ with computational basis states $|0\rangle$ and $|1\rangle$; a single-qubit state $|\psi_2\rangle$ is a linear combination of $|0\rangle$ and $|1\rangle$ with amplitudes satisfying the normalization condition. Quantum gates are unitary matrices; single-qubit gates are therefore $2\times 2$ unitary matrices. Typical single-qubit gates include the Pauli gates $Z$ and $X$,
\begin{equation}
	\label{pauli x z}
	Z=\begin{pmatrix}1&0\\0&-1\end{pmatrix},\quad X=\begin{pmatrix}0&1\\1&0\end{pmatrix},
\end{equation}
the identity gate $1\!\!1_2 = Z^2$, and the Hadamard gate $H=\frac{1}{\sqrt{2}}(X+Z)$.

A two-qubit state space $\mathscr{H}_2 \otimes \mathscr{H}_2$ has, in addition to the product basis $|00\rangle$, $|01\rangle$, $|10\rangle$, and $|11\rangle$, the Bell basis as an orthonormal basis. The Bell basis states are precisely the maximally entangled bipartite pure states. We consider $n$ pairs of Bell states $|\phi(\alpha_k\beta_k)\rangle$ for $k=1,\dots, n$, with $\alpha_k, \beta_k \in \{0,1\}$.

Given the Bell state $|\phi\rangle\equiv|\phi(00)\rangle$,
\begin{equation}
	\label{twoqubitbellstate00}  
	|\phi(00)\rangle=\frac{1}{\sqrt{2}}(|00\rangle+|11\rangle),
\end{equation}
the four Bell basis states have a unified form  
\begin{equation}
	\label{twoqubitbellstate}  
	|\phi(\alpha_k\beta_k)\rangle=(T(\alpha_k\beta_k)\otimes 1\!\!1_2)|\phi\rangle,
\end{equation}
with $T(\alpha_k\beta_k)=Z^{\alpha_k} X^{\beta_k}$, so that they can be created on a quantum circuit:
\begin{equation}
	|\phi(\alpha_k\beta_k)\rangle=CNOT \cdot (H\otimes 1\!\!1_2)|\alpha_k\beta_k\rangle,
\end{equation}
where the CNOT gate is
\begin{equation}
	CNOT=|0\rangle\langle 0| \otimes 1\!\!1_2 + |1\rangle\langle 1| \otimes X.
\end{equation} 

As an orthonormal basis of $\mathscr{H}_2 \otimes \mathscr{H}_2$, the Bell basis states $|\phi(\alpha_k\beta_k)\rangle$ satisfy the orthonormality condition:
\begin{equation}
	\label{two-qubit bell orthonormal condition}
	\langle\phi(\alpha_k^\prime\beta_k^\prime)|\phi(\alpha_k\beta_k)\rangle= \delta_{\alpha^\prime_k\alpha_k}\delta_{\beta_k^\prime\beta_k},
\end{equation}
with the Kronecker delta $\delta$. This condition is equivalent to 
\begin{equation}
	\frac {1} {2} 	\operatorname{tr}\bigl( T^\dagger(\alpha_k^\prime\beta_k^\prime) T(\alpha_k\beta_k) \bigr)
	=\delta_{\alpha_k^\prime\alpha_k}
	\delta_{\beta_k^\prime\beta_k},
\end{equation}
with $\dagger$ denoting Hermitian conjugation. 

The completeness relation for the orthonormal basis $|\phi(\alpha_k\beta_k)\rangle$ is expressed as 
\begin{equation}
	\label{two-qubit bell completeness relation}
	\sum_{\alpha_k,\beta_k=0}^1 	|\phi(\alpha_k\beta_k)\rangle\langle\phi(\alpha_k\beta_k)|
	=1\!\!1_2^{\otimes 2} 
\end{equation} 
with the identity operator $1\!\!1_2^{\otimes 2}$. It can be derived from the simpler equation
\begin{equation}
	\frac {1} {2} \sum_{\alpha_k,\beta_k=0}^1 
	T(\alpha_k\beta_k)	|i_k\rangle\langle j_k| T^\dagger(\alpha_k\beta_k)
	=  \delta_{i_k j_k}  1\!\!1_2.
\end{equation}

There are two commuting observables for which the Bell basis states form a set of eigenvectors. In terms of the projectors \(|\phi(\alpha_k\beta_k)\rangle\langle\phi(\alpha_k\beta_k)|\), they have the following spectral decompositions:
\begin{equation}
	\label{phase-parity-observable}
	\begin{aligned}
		&X\otimes X=\sum_{\alpha_k,\beta_k=0}^1 (-1)^{\alpha_k} |\phi(\alpha_k\beta_k)\rangle \langle\phi(\alpha_k\beta_k)|,\\
		&Z\otimes Z=\sum_{\alpha_k,\beta_k=0}^1 (-1)^{\beta_k} |\phi(\alpha_k\beta_k)\rangle \langle\phi(\alpha_k\beta_k)|,
	\end{aligned}
\end{equation}
where \(\alpha_k\) is called the phase-bit, \(X\otimes X\) the phase-bit observable, \(\beta_k\) the parity-bit, and \(Z\otimes Z\) the parity-bit observable. Note that the Bell measurement can be viewed either directly as measurements of these two observables or simply as the projective measurement specified by the projectors \(|\phi(\alpha_k\beta_k)\rangle\langle\phi(\alpha_k\beta_k)|\).

The Bell states and the Bell measurement are key ingredients in the standard description of quantum teleportation \cite{Bennett_1993}. This protocol enables an unknown qubit \(|\psi_2\rangle\) to be sent from Alice to Bob without using a quantum communication channel. Alice and Bob share the Bell state \(|\phi(00)\rangle\), and the prepared state satisfies the following equality:
\begin{equation}\label{teleportation equation 2}
	|\psi_2\rangle\otimes|\phi(00)\rangle
	=\frac{1}{2}\sum_{\alpha_k,\beta_k=0}^1|\phi(\alpha_k\beta_k)
	\rangle\otimes T^\dagger(\alpha_k\beta_k)|\psi_2\rangle,
\end{equation}
which was originally called the teleportation equation in \cite{Zhang2006,Zhang2009}. Alice then performs the Bell measurement on her two qubits and informs Bob of her measurement outcomes—the two bits \((\alpha_k,\beta_k)\)—through a classical channel. Finally, Bob applies the unitary correction operators \(T(\alpha_k\beta_k)\) to his qubit to recover the unknown qubit \(|\psi_2\rangle\).

\section{Generalized two-qudit Bell states}

Some important algebraic properties of generalized two-qudit Bell states are described throughout this paper. A generalized two-qudit Bell basis is obtained by applying a set of $d\times d$ unitary matrices to a specified generalized two-qudit Bell state, which constitutes an orthonormal basis for a $d^2$-dimensional Hilbert space. The basis theorem and the basis group are introduced and investigated; they are developed from the observation that the local action of a $d \times d$ unitary matrix on each generalized two-qudit Bell basis state induces another orthonormal basis.

\subsection{Basic properties}

Let $\{|i\rangle\}_{i=0}^{d-1}$ be an orthonormal basis for the $d$-dimensional Hilbert space $\mathscr{H}_d$. A qudit state $|\psi_d\rangle$ is a linear combination of the basis states $|i\rangle$ with complex amplitudes satisfying the normalization condition. Define the generalized two-qudit Bell state $|\Omega\rangle$ via the Schmidt decomposition
\begin{equation}
	\label{twoquditbellstate}
	|\Omega\rangle \equiv\frac{1}{\sqrt{d}}\sum_{i=0}^{d-1}|i\rangle\otimes|i\rangle,
\end{equation}
which has the following algebraic properties.

First, let $M$ be a linear operator on $\mathscr{H}_d$. Its local action on the generalized Bell state $|\Omega\rangle$ satisfies
\begin{equation}\label{12}
	({M}\otimes 1\!\!1_d)|\Omega\rangle=(1\!\!1_d\otimes{M}^\intercal)|\Omega\rangle
\end{equation}
with the $d\times d$ identity matrix $1\!\!1_d$ and the matrix transpose $\intercal$. That is, when the action of $M$ moves from the first qudit Hilbert space to the second, it is transformed via the matrix transpose. Note that $M$ is not required to be unitary.

Write a $d\times d$ matrix $M$ as a product of two square matrices $L$ and the transpose of $N$, namely $M=L N^\intercal$. The local action of $M$ on $|\Omega\rangle$ is then
\begin{equation}
	\begin{aligned}
		|\mathcal{M}\rangle
		& \equiv (M\otimes1\!\!1_d)|\Omega\rangle\\
		&=(L\otimes N)|\Omega\rangle.
	\end{aligned}
\end{equation}
When $M, L,$ and $N$ are unitary matrices, the state $|\mathcal{M}\rangle$ has the general Schmidt decomposition form of a two-qudit pure state. Similarly, write
$$|\mathcal{M}^\intercal\rangle \equiv  (M^\intercal\otimes1\!\!1_d)|\Omega\rangle$$
for the local action of the transpose of $M$ on $|\Omega\rangle$.

Second, the inner product between two generalized Bell states generated by the local action of two $d\times d$ matrices $L$ and $N$ is given by the trace of a matrix product,
\begin{equation}\label{13}
	\langle\Omega|({N}^\dagger\otimes 1\!\!1_d)({L}\otimes 1\!\!1_d)|\Omega\rangle=\frac{1}{d}
	\operatorname{tr}\bigl({N}^\dagger {L} \bigr).
\end{equation}	
In particular, when $N$ is the identity, this reduces to the trace of $L$ divided by $d$. When both $N$ and $L$ are the identity matrix, it gives the normalization condition of $|\Omega\rangle$. 

Third, with subscripts $C, A,$ and $B$ denoting three distinct single-qudit Hilbert spaces, respectively, the transfer operator $\mathcal{T}_{CB}\equiv\sum_i |i\rangle_B \, {}_C\langle i|$ maps a qudit state $|\psi_d\rangle$ from $\mathscr{H}_C$ to $\mathscr{H}_B$:
\begin{equation}
	\label{transfer operator1}
	\mathcal{T}_{CB}|\psi_d\rangle_C=|\psi_d\rangle_B.
\end{equation}
This operator can be expressed as an inner product between generalized two-qudit Bell states and involves three qudits $C, A,$ and $B$:
\begin{equation}\label{transfer operator2}
	_{CA}\langle\Omega|\Omega\rangle_{AB}=\frac{1}{d}\mathcal{T}_{CB},
\end{equation}
so that the following equality holds:
\begin{equation}
	({_{CA}\langle}\Omega|\otimes1\!\!1_d)(|\psi_d\rangle_C\otimes|\Omega\rangle_{AB})
	=\frac{1}{d}|\psi_d\rangle_B,
\end{equation}
whose special case for $d=2$ is equivalent to the teleportation equation \eqref{teleportation equation 2}. Consequently, the transfer operator plays a crucial role in the study of quantum teleportation.

\subsection{Generalized two-qudit Bell bases}

Now we formulate a set of generalized two-qudit Bell states as 
\begin{equation}
	|\Omega(a)\rangle=\frac{1}{\sqrt{d}}\sum_{i=0}^{d-1}U_a|i\rangle\otimes|i\rangle,
\end{equation}
with a set of $d\times d$ unitary matrices $U_a$, for $a=0,1,\dots,d^2-1$, and the identity matrix 
$U_0=1\!\!1_d$. Note that $|\Omega(0)\rangle\equiv |\Omega\rangle$. All these generalized
Bell states are maximally entangled bipartite pure states. 

Consider the case in which the set $|\Omega(a)\rangle$ forms an orthonormal basis for the two-qudit Hilbert space $\mathscr{H}_d\otimes\mathscr{H}_d$. Such states must then satisfy the orthonormality condition
$\langle \Omega(a)|\Omega(b)\rangle =\delta_{ab}$ and the completeness relation
\begin{equation}
	\sum_{a=0}^{d^2-1}|\Omega(a)\rangle\langle\Omega(a)|=1\!\!1_{d^2} 
\end{equation}
with the $d^2\times d^2$ identity matrix $1\!\!1_{d^2}$. 

The orthonormality condition leads to constraints on the unitary matrices $U_a$ and $U_b$ in the Hilbert--Schmidt inner product \cite{NC2011}:
\begin{equation}
	\label{orthonormal2}
	\frac{1}{d}\operatorname{tr}\bigl(U_a^\dagger U_b\bigr)=\delta_{ab}.
\end{equation}
That is, the set $\{U_a\}$ forms an orthonormal basis for the space of all $d\times d$ matrices. The completeness relation can be derived from the simplified equation
\begin{equation}
	\label{completeness2}
	\frac{1}{d}\sum_{a=0}^{d^2-1}U_a |i\rangle\langle j|U_a^\dagger=\delta_{ij} 1\!\!1_d,
\end{equation}
called the reduced completeness relation in this paper, with its component form
\begin{equation}
	\label{completeness3}
	\frac{1}{d}\sum_{a=0}^{d^2-1} (U_a)_{li}  (U_a^\dagger)_{jm}=\delta_{lm}\delta_{ij}.
\end{equation}

It is straightforward to construct further orthonormal bases from the basis $|\Omega(a)\rangle$. For example, given a $d^2 \times d^2$ unitary matrix $U$, the set $\{U|\Omega(a)\rangle\}$ is also an orthonormal basis. Since the transpose of $U_a$, namely $U_a^\intercal$, is unitary, the states $|\Omega^\intercal(a)\rangle=(U_a^\intercal \otimes 1\!\!1_d)|\Omega \rangle$ form another orthonormal basis.

In a $d^2$-dimensional Hilbert space, when the number of orthonormal states $|\Omega(a)\rangle$ is $d^2$, the completeness relation follows from the orthonormality condition; thus the two are not independent. The projectors $|\Omega(a)\rangle\langle\Omega(a)|$ play important roles in projective measurement theory for quantum teleportation and in the spectral decomposition of observables that have generalized Bell states as eigenvectors. Therefore, in this paper, we focus on the orthonormality condition while also studying the completeness relation.

\subsection{The basis theorem and basis group}
\label{basisgroup} 

For an orthonormal basis state $|\Omega(a)\rangle$, two new states can be defined by left-multiplying $U_a$ by a $d\times d$ matrix $M$,
\begin{equation}
	|M\Omega(a)\rangle\equiv (M U_a\otimes 1\!\!1_d)|\Omega\rangle,
\end{equation}
where $|M\Omega(0)\rangle\equiv |\mathcal{M}\rangle$, and by right-multiplying $U_a$ by $M$,
\begin{equation}
	|\Omega M(a)\rangle \equiv (U_a M\otimes 1\!\!1_d)|\Omega\rangle,
\end{equation}
where $|\Omega M(0)\rangle \equiv |\mathcal{M}\rangle$. Similarly, denote
$|\Omega M^\intercal(a)\rangle=(U_a \otimes M)|\Omega\rangle$.
With the basis $|\Omega(a)\rangle$ and the matrix $M$, we propose the so-called basis theorem. 

\begin{theorem}
	\label{basislemma}
	Given the orthonormal basis $|\Omega(a)\rangle$, the set of states $|M\Omega(a)\rangle$ (or $|\Omega M(a)\rangle$) forms an orthonormal basis for the Hilbert space $\mathscr{H}_d\otimes\mathscr{H}_d$ if and only if the matrix $M$ is a $d\times d$ unitary matrix. 
\end{theorem}

When $M$ is unitary, the proof of the basis theorem is straightforward. First, both sets $|\Omega(a)\rangle$ and $|\Omega^\intercal(a)\rangle$ are orthonormal bases. Second, $M\otimes 1\!\!1_d$ and $M^\intercal\otimes 1\!\!1_d$ are unitary matrices, so the sets $|M\Omega(a)\rangle$ and $(M^\intercal\otimes 1\!\!1_d)|\Omega^\intercal(a)\rangle$ form orthonormal bases. Third, since $M^\intercal U_a^\intercal =(U_a M)^\intercal$, it follows that the set $|\Omega M(a)\rangle$ is also an orthonormal basis.   

To verify that the orthonormal basis $|M\Omega(a)\rangle$ (or $|\Omega M(a)\rangle$) implies the unitarity of $M$, it is best to introduce the concept of a basis group. A set of matrices is called a basis group if and only if it forms a group and its elements satisfy the Hilbert--Schmidt orthonormality condition. A basis group is called a unitary basis group if and only if its elements are unitary matrices. A basis group is called a special basis group if and only if the conjugate transpose of each element is also in the set. For convenience, we use only the term ``basis group" in the subsequent discussion, even when unitary or special basis groups are involved. This is justified because the basis group in this paper is generated by either Pauli matrices or generalized Pauli matrices; hence its elements are products of such matrices multiplied by global phase factors.  

Consider the case where the states $|M\Omega(a)\rangle$ form an orthonormal basis. Then one has the Hilbert--Schmidt inner product
\begin{equation}
	\frac{1}{d}\operatorname{tr}\bigl(M^\dagger M U_b U_a^\dagger \bigr)=\delta_{ab}.
\end{equation}
Since the set $\{U_a\}$ is a basis group, as previously described, the products $U_b U_a^\dagger$ are elements of this basis group, so that the set of all possible $U_b U_a^\dagger$ is an orthonormal basis for the space of $d\times d$ matrices, just like the set $\{U_a\}$. Consequently, $M^\dagger M$ must be the identity matrix, because the expansion coefficients of $M^\dagger M$ with respect to this orthonormal basis are precisely the Kronecker delta. Alternatively, the orthonormal basis $|\Omega M(a)\rangle$ gives
\begin{equation}
	\frac{1}{d}\operatorname{tr}\bigl( M M^\dagger U_a^\dagger U_b \bigr)=\delta_{ab},
\end{equation}
which implies that $M M^\dagger$ is also the identity. Therefore, $M$ is unitary.

Beyond the proof using the orthonormality condition, the completeness relation for the orthonormal basis $|M\Omega(a)\rangle$ also implies the unitarity of $M$. A direct algebraic calculation yields the reduced completeness relation for $|\Omega M(a)\rangle$:
\begin{equation}
	\frac{1}{d}\sum_{a=0}^{d^2-1}U_a M |i\rangle\langle j|M^\dagger  U_a^\dagger=(M^\dagger M)_{ji}  1\!\!1_d.
\end{equation}
This shows that the orthonormal basis $|\Omega M(a)\rangle$ enforces the unitarity of $M$.  

The basis theorem is indeed a generalization of Kauffman's basis lemma \cite{Kauffman2005}, which he briefly described without a mathematical proof and solely in the context of generalized multiple qubit states. The basis theorem applies to generalized Bell states under the local action of $M$, specifically the generalized two-qudit Bell states. We note that the connection between generalized two-qudit Bell states and generalized multiple qubit states will be clarified later in this paper.

\section{Generalized Bell states via generalized Pauli matrices}

A typical example of generalized two-qudit Bell states \(|\Omega(a)\rangle\) is obtained by choosing products of generalized Pauli matrices \cite{DM2025, RDF2003, PZ1988} as the local unitary operators \(U_a\). First, we briefly introduce the generalized Pauli matrices and their algebraic properties. Second, we present the associated generalized two-qudit Bell states and discuss the basis group and basis theorem. Third, we construct in detail a complete set of observables for which these generalized two-qudit Bell states are eigenvectors.

\subsection{Generalized Pauli matrices}

The generalized Pauli matrices \(\mathcal{X}\) and \(\mathcal{Z}\) \cite{DM2025, RDF2003, PZ1988} are respectively defined by 
\begin{equation}
	\label{general pauli x}
	\mathcal{X}|i\rangle \equiv |i\oplus 1\rangle,
\end{equation}
with addition modulo \(d\), i.e., \(i\oplus d \equiv i \pmod d\), for \(i=0,\dots,d-1\), and 
\begin{equation}
	\label{general pauli z}
	\mathcal{Z}|i\rangle  \equiv   \omega^i| i\rangle,
\end{equation}
with the complex number \(\omega=e^{2\pi i/d}\) satisfying
\begin{equation}\label{omega property}
	\sum_{i=0}^{d-1}\omega^{ki}
	=d\,\delta_{k,nd},
\end{equation}
where \(k\) and \(n\) are integers. These matrices are clearly unitary and have finite order \(d\):
\begin{equation}
	\mathcal{X}^d=1\!\! 1_d, \quad \mathcal{Z}^d=1\!\! 1_d,
\end{equation}
and they satisfy the commutation relation
\begin{equation}
	\mathcal{Z}\mathcal{X}=\omega\,\mathcal{X}\mathcal{Z}.
\end{equation}
Note that they are not Hermitian except for the case \(d=2\).

\subsection{Generalized two-qudit Bell basis, the basis group, and the basis theorem}

Let $U_{\alpha\beta}=\mathcal{Z}^\alpha\mathcal{X}^\beta$ with 
$\alpha,\beta=0,1,\dots,d-1$. It is straightforward to verify that the $U_{\alpha\beta}$ satisfy the orthonormality condition
\begin{equation}
	\frac{1}{d} \operatorname{tr}\bigl(U^\dagger_{\alpha^\prime\beta^\prime} U_{\alpha\beta }\bigr)
	=\delta_{\alpha^\prime \alpha}\delta_{\beta^\prime \beta},
\end{equation}
and the reduced completeness relation
\begin{equation}
	\frac{1}{d}\sum_{\alpha,\beta=0}^{d-1} U_{\alpha\beta} |i\rangle\langle j| U_{\alpha\beta}^\dagger=\delta_{ij} 1\!\!1_d.
\end{equation}
Thus, the generalized two-qudit Bell states constructed via generalized Pauli matrices are
\begin{equation}\label{qudit GBS}
	|\Omega(\alpha\beta)\rangle
	=(\mathcal{Z}^\alpha\mathcal{X}^\beta\otimes1\!\!1_d)|\Omega\rangle.
\end{equation}
The relevant algebraic calculation for the orthonormality condition is
\begin{equation}
	\begin{aligned}
		\langle\Omega(\alpha^\prime\beta^\prime)|\Omega(\alpha\beta)\rangle
		&=\frac{1}{d}\sum_{i=0}^{d-1}\omega^{\alpha(i+\beta)}{\omega}^{-\alpha^\prime(i+\beta^\prime)}\delta_{\beta\beta^\prime}\\
		&=\delta_{\alpha^\prime \alpha}\delta_{\beta^\prime \beta},
	\end{aligned}
\end{equation}
and for the completeness relation it is
\begin{equation}
	\begin{aligned}
		\sum_{\alpha,\beta=0}^{d-1} |\Omega(\alpha\beta)\rangle \langle\Omega(\alpha\beta)|
		&=\frac{1}{d}\sum_{\alpha,\beta,i,j=0}^{d-1}\omega^{\alpha(i+\beta)}{\omega}^{-\alpha(j+\beta)}|{i+\beta}\rangle\langle {j+\beta}|\otimes|i\rangle\langle j|\\ 
		&=1\!\!1_d\otimes1\!\!1_d.
	\end{aligned}
\end{equation}
Hence, the set $\{|\Omega(\alpha\beta)\rangle\}$ is an orthonormal basis for the two-qudit Hilbert space $\mathscr{H}_d\otimes\mathscr{H}_d$.

Consider the unitary matrices $\omega^\gamma U_{\alpha\beta}$ for $\gamma=0,1,\dots,d-1$. They satisfy
\begin{equation}
	(\omega^\gamma U_{\alpha,\beta})^\dagger=\omega^{(d-1)\gamma+(d-1)^3\alpha\beta} U_{(d-1)\alpha,(d-1)\beta},
\end{equation}
and thus the set of such matrices forms a basis group as defined in Subsec.~\ref{basisgroup}.
Given a $d\times d$ unitary matrix $M$, the two extended orthonormal bases are respectively constructed as 
$|M\Omega (\alpha\beta)\rangle=(M \otimes 1\!\!1_d)|\Omega(\alpha\beta)\rangle$
and 
\begin{equation}
	|\Omega M(\alpha\beta)\rangle=(U_{\alpha\beta} M\otimes 1\!\!1_d)|\Omega\rangle.
\end{equation}
As a result, the basis theorem for generalized two-qudit Bell states using generalized Pauli matrices can be stated in the following corollary.   
\begin{corollary}
	Let $M$ be a linear operator on a single-qudit Hilbert space $\mathscr{H}_d$. Then the set $\{|M\Omega(\alpha\beta)\rangle\}$ (or $\{|\Omega M(\alpha\beta)\rangle\}$) forms an orthonormal basis for the two-qudit Hilbert space $\mathscr{H}_d\otimes \mathscr{H}_d$ if and only if $M$ is unitary.
\end{corollary}

\subsection{Observables for generalized two-qudit Bell states}

Let us construct a complete set of commuting observables that define the generalized two-qudit Bell states \(|\Omega(\alpha\beta)\rangle\) as common eigenvectors. Two commuting operators,
\begin{equation}
	A(k)\equiv \mathcal{X}^k\otimes \mathcal{X}^k, \qquad 
	B(k)\equiv \mathcal{Z}^k\otimes(\mathcal{Z}^\dagger)^k,
\end{equation}
have \(|\Omega(\alpha\beta)\rangle\) as common eigenvectors, and their eigenequations take the form 
\begin{equation}
	A(k)|\Omega(\alpha\beta)\rangle
	=\omega^{-k\alpha}|\Omega(\alpha\beta)\rangle,\qquad   B(k)|\Omega(\alpha\beta)\rangle
	=\omega^{k\beta}|\Omega(\alpha\beta)\rangle,
\end{equation}
where \(k=0,1,\dots,d-1\). The Hermitian conjugates \(A^\dagger(k)\) and \(B^\dagger(k)\) also have 
\(|\Omega(\alpha\beta)\rangle\) as common eigenvectors:
\begin{equation}
	A^\dagger(k) |\Omega(\alpha\beta)\rangle
	=\omega^{k\alpha}|\Omega(\alpha\beta)\rangle,\qquad   B^\dagger(k) |\Omega(\alpha\beta)\rangle
	=\omega^{-k\beta}|\Omega(\alpha\beta)\rangle.
\end{equation}

It turns out that the following four Hermitian operators,
\begin{equation}
	\begin{aligned}
		&OX_+(k)=\frac{1}{2}\bigl(A(k) + A^\dagger(k)\bigr),\\
		&OX_-(k)=\frac{i}{2}\bigl(A(k) - A^\dagger(k)\bigr),\\
		&OZ_+(k)=\frac{1}{2}\bigl(B(k) + B^\dagger(k)\bigr),\\
		&OZ_-(k)=-\frac{i}{2}\bigl(B(k) - B^\dagger(k)\bigr),
	\end{aligned}
\end{equation}
have \(|\Omega(\alpha\beta)\rangle\) as common eigenvectors, with the respective eigenvalues
\begin{equation}\label{O for n qudit}
	\begin{aligned}
		&OX_+(k)|\Omega(\alpha\beta)\rangle=\cos\frac{2k\alpha\pi}{d}\,|\Omega(\alpha\beta)\rangle,\\
		&OX_-(k)|\Omega(\alpha\beta)\rangle=\sin\frac{2k\alpha\pi}{d}\,|\Omega(\alpha\beta)\rangle,\\
		&OZ_+(k)|\Omega(\alpha\beta)\rangle=\cos\frac{2k\beta\pi}{d}\,|\Omega(\alpha\beta)\rangle,\\
		&OZ_-(k)|\Omega(\alpha\beta)\rangle=\sin\frac{2k\beta\pi}{d}\,|\Omega(\alpha\beta)\rangle.
	\end{aligned}
\end{equation}
For two-qubit Bell states, i.e., for \(d=2\), the observables \(OX_+(1)\) and \(OZ_+(1)\) coincide with the phase-bit observable \(X\otimes X\) and the parity-bit observable \(Z\otimes Z\) in (\ref{phase-parity-observable}), while \(OX_-(1)\) and \(OZ_-(1)\) reduce to the zero operator. Note that for \(k\neq 0\), the above set of four observables is sufficient to uniquely determine the values of \(\alpha\) and \(\beta\); hence it constitutes a complete set of observables defining the generalized two-qudit Bell states \(|\Omega(\alpha\beta)\rangle\) as unique eigenvectors.

A complete set of observables defining the states \(|M\Omega(\alpha\beta)\rangle\) (or \(|\Omega M(\alpha\beta)\rangle\)) can be obtained by applying the unitary transformation \(M\otimes 1\!\! 1_d\) 
(or \(1\!\! 1_d \otimes M^{\intercal}\)) to the complete set of observables for \(|\Omega(\alpha\beta)\rangle\). For example, the eigenequation involving the observable, its eigenvalue, and the eigenvector \(|M\Omega(\alpha\beta)\rangle\) reads
\begin{equation}\label{M_OX+k}
	(M\otimes 1\!\! 1_d)	OX_+(k)  (M^\dagger \otimes 1\!\! 1_d)  |M\Omega(\alpha\beta)\rangle=\cos\frac{2k\alpha\pi}{d}\,|M\Omega(\alpha\beta)\rangle.
\end{equation}
Similarly, for \(|\Omega M(\alpha\beta)\rangle\), the corresponding eigenequation has the form
\begin{equation}\label{M_OZ+k}
	(1\!\!  1_d\otimes M^{\intercal})	OZ_+(k)  (1\!\! 1_d\otimes M^{\ast})  |\Omega M(\alpha\beta)\rangle=\cos\frac{2k\beta\pi}{d}\,|\Omega M(\alpha\beta)\rangle,
\end{equation}
where \(\ast\) denotes complex conjugation.

\section{Generalized multiple qubit Bell states}

A generalized multiple qubit Bell state is referred to as a generalized $2n$-qubit Bell state in this paper. It is an $n$-fold tensor product of two-qubit Bell states, optionally with a twist operation, where the twist operator is expressed as a product of SWAP (or permutation) gates. Similar to the previous construction of a generalized two-qudit Bell basis, a generalized $2n$-qubit Bell basis—as an orthonormal basis for a $2^{2n}$-dimensional Hilbert space—is formulated by the local action of $n$-fold tensor products of Pauli gates on a given generalized $2n$-qubit Bell state. Each basis state can be created by a quantum circuit consisting of Hadamard, CNOT, and SWAP gates.

Furthermore, the basis group for a generalized $2n$-qubit Bell basis is a set of $n$-fold tensor products of Pauli gates, and the associated basis theorem can be verified from the orthonormality condition or the completeness relation. In particular, the proof of the basis theorem can be carried out using the Hilbert--Schmidt inner product or via a detailed study of the algebraic properties of tensor products of Pauli gates presented in Appendix A. Moreover, in addition to a complete set of observables that have the generalized $2n$-qubit Bell states as eigenvectors, we propose and compute the generalized concurrence of a \(2n\)-qubit state with respect to the generalized \(2n\)-qubit Bell basis.

\subsection{Twist of tensor products of Bell states}

Consider a $2^{2n}$-dimensional Hilbert space, namely a $2n$-qubit Hilbert space.
The simplest construction of a generalized $2n$-qubit Bell state $|\Phi_{2n}\rangle$ is the $n$-fold tensor product of two-qubit Bell states $|\phi\rangle$ from (\ref{twoqubitbellstate}), so that
\begin{equation}
	|\Phi_{2n}\rangle \equiv |\phi\rangle^{\otimes n}.
\end{equation}    
Another generalized $2n$-qubit Bell state can be derived from the two-qudit Bell state $|\Omega\rangle$ in (\ref{twoquditbellstate}) at $d=2^n$:
\begin{equation}
	|\mathcal{B}_{2n}\rangle \equiv  |\Omega\rangle|_{d=2^n}.
\end{equation}  
Hence there exists a twist operator $\tau_{2n}$ between $|\Phi_{2n}\rangle$ and $|\mathcal{B}_{2n}\rangle$,
\begin{equation}
	|\mathcal{B}_{2n}\rangle=\tau_{2n}|\Phi_{2n}\rangle,
\end{equation}  
and it is useful to study the properties of $\tau_{2n}$ in a little more detail.

First, a one-to-one correspondence between a decimal number $i$ and an $n$-bit string $i_1i_2\cdots i_n$ is established as
\begin{equation}
	i=\sum_{k=1}^n i_k 2^{n-k}.
\end{equation}
Note that we use the convention $\underline{i} \equiv i_1 i_2 \dots i_n$ throughout this paper. A state $|\underline{i}\rangle$ in the $2^n$-dimensional Hilbert space is then expressed as a tensor product of states in the two-dimensional Hilbert space:
\begin{equation}\label{2 and 10}
	|\underline{i}\rangle \equiv  |i_1\rangle\otimes|i_2\rangle\otimes\cdots\otimes|i_n\rangle,
\end{equation}  
with the simplified notation $|\underline{i}\rangle \equiv |i_1 i_2\cdots i_n\rangle$. 

Similarly, another decimal number $j$ has the representation of an $n$-bit string $j_1j_2\cdots j_n$. For $\underline{i}$ and $\underline{j}$, we respectively define two types of $2n$-bit strings:
\[
\underline{i}\,\underline{j} \equiv  i_1\cdots i_n j_1\cdots j_n, \qquad
\underline{ij} \equiv   i_1j_1\cdots i_n j_n,
\]
so that the corresponding $2n$-qubit quantum states are constructed as
\begin{equation}
	|\underline{i}\underline{j} \rangle  \equiv   |i_1 \cdots i_n\rangle|j_1 \cdots j_n\rangle,
\end{equation}
and
\begin{equation}
	|\underline{ij} \rangle \equiv   |i_1 j_1\rangle\cdots |i_n j_n\rangle.
\end{equation} 
These define the twist operator $\tau_{2n}$ via
\begin{equation}
	|\underline{i}\,\underline{j} \rangle=\tau_{2n} 	|\underline{ij} \rangle,
\end{equation}
so that $\tau_{2n}$ has the explicit form
\begin{equation}
	\tau_{2n}=\sum_{i_1,\dots,i_n=0}^1\sum_{j_1,\dots,j_n=0}^1 
	|i_1 \cdots i_n j_1 \cdots j_n\rangle  \langle i_1 j_1\cdots i_n j_n |.
\end{equation}

It follows that $\tau_{2n}$ can be expressed as a product of a sequence of permutations that transform
a $2n$-qubit string labeled by $i_1 j_1\cdots i_n j_n$ to another $2n$-qubit string labeled by $i_1 \cdots i_n j_1 \cdots j_n$, namely
\begin{equation}
	\tau_{2n}=M^{n+1}_2 M^{n+2}_4 \cdots M^{2n-1}_{2n-2},
\end{equation}
where each $M^{n+k}_{2k}$ is a product of adjacent transpositions on neighboring qubits:
\begin{equation}
	M^{n+k}_{2k}=P^{n+k}_{n+k-1} P^{n+k-1}_{n+k-2} \cdots P^{2k+1}_{2k},
\end{equation}
with $P^{m+1}_m$ denoting the SWAP gate that interchanges the $m$th and $(m+1)$th qubits:
\begin{equation}
	SWAP |m\rangle |m+1\rangle =  |m+1\rangle |m\rangle.
\end{equation}
Thus $M^{n+k}_{2k}$ moves the $2k$th qubit $j_k$ in the $2n$-qubit string
labeled by 
\begin{equation}
	i_1 j_1 \cdots i_k j_k i_{k+1} i_{k+2}\cdots i_n j_{k+1} j_{k+2} \cdots j_n 
\end{equation}   
across the substring $i_{k+1} i_{k+2}\cdots i_n$ to obtain the new $2n$-qubit string
labeled by 
\begin{equation}
	i_1 j_1 \cdots i_{k-1} j_{k-1} i_k i_{k+1} i_{k+2}\cdots i_n j_{k} j_{k+1} j_{k+2} \cdots j_n. 
\end{equation}   

Consequently, the quantum circuit that implements $\tau_{2n}$ consists of a sequence of SWAP gates. For example,
$\tau_2$ is the identity gate $\tau_2= 1\!\!1_2$, and $\tau_4=M_2^3$ is realized as
\begin{equation}
	\tau_4=1\!\!1_2 \otimes SWAP\otimes 1\!\!1_2. 
\end{equation}  
The total number of SWAP gates required to implement $\tau_{2n}$ is at least $\frac{n(n-1)}{2}$, which corresponds to at least $\frac{3n(n-1)}{2}$ CNOT gates.

\subsection{Generalized $2n$-qubit Bell bases}

A generalized $2n$-qubit Bell basis is an orthonormal basis for a $2n$-qubit Hilbert space. The first construction is the set of $n$-fold tensor products of two-qubit Bell states,
\begin{equation}
	|\Phi_{2n}(\underline{\alpha\beta})\rangle
	=|\phi(\alpha_1\beta_1)\rangle \otimes \cdots \otimes |\phi(\alpha_n\beta_n)\rangle, 
\end{equation}    
and has the conventional tensor product representation 
\begin{equation}
	|\Phi_{2n}(\underline{\alpha\beta})\rangle
	\equiv \bigotimes_{k=1}^n |\phi(\alpha_k\beta_k)\rangle, 
\end{equation}   
with the $2n$-bit string 
\[\underline{\alpha\beta} \equiv \alpha_1\beta_1\alpha_2\beta_2\cdots \alpha_n\beta_n\] generated by the two $n$-bit strings $\underline{\alpha} \equiv  \alpha_1\alpha_2\cdots\alpha_n$ and $\underline{\beta} \equiv \beta_1\beta_2\cdots \beta_n$. In terms of $2n$-fold tensor products of Pauli gates,
\begin{equation}
	\bigotimes_{k=1}^n ( T(\alpha_k\beta_k)\otimes 1\!\!1_2)
	\equiv  T(\alpha_1\beta_1)\otimes 1\!\!1_2 \otimes \cdots
	\otimes T(\alpha_n\beta_n)\otimes 1\!\!1_2,
\end{equation}   
thus \(|\Phi_{2n}(\underline{\alpha\beta})\rangle\) has the equivalent form
\begin{equation}
	|\Phi_{2n}(\underline{\alpha\beta})\rangle
	=\bigotimes_{k=1}^n ( T(\alpha_k\beta_k)\otimes 1\!\!1_2)|\Phi_{2n}\rangle.
\end{equation}   

The second construction is the set of generalized two-qudit Bell states, with each qudit representing $n$ qubits,
\begin{equation}
	|\mathcal{B}_{2n}(\underline{\alpha\beta})\rangle
	=(T_n(\underline{\alpha\beta})\otimes1\!\!1_2^{\otimes n})|\mathcal{B}_{2n}\rangle,
\end{equation}
where the $n$-fold tensor product of Pauli gates, 
\begin{equation}
	T_n(\underline{\alpha\beta})=T(\alpha_1\beta_1)\otimes T(\alpha_2\beta_2)\otimes\cdots\otimes T(\alpha_n\beta_n),
\end{equation}
uses the simplified notation
\begin{equation}
	T_n(\underline{\alpha\beta}) \equiv   \bigotimes_{k=1}^n Z^{\alpha_k}X^{\beta_k}.
\end{equation}
With the twist operator $\tau_{2n}$, the relation between the two types of generalized Bell bases 
is given by
\begin{equation}
	|\mathcal{B}_{2n}(\underline{\alpha\beta})\rangle
	=\tau_{2n}|\Phi_{2n}(\underline{\alpha\beta})\rangle,
\end{equation}
where the equality 
\begin{equation}
	\tau_{2n}\bigotimes_{k=1}^n ( T(\alpha_k\beta_k)\otimes 1\!\!1_2)
	=(T_n(\underline{\alpha\beta})\otimes 1\!\!1_2^{\otimes n} )\tau_{2n}
\end{equation}
has already been used. Note that $\tau_{2n}$ has the crucial property
\begin{equation}
	\tau_{2n}\left(\bigotimes_{i=1}^{2n} O_i \right)	\tau_{2n}^\dagger 
	=\bigotimes_{j=1}^{n} O_{2j-1}  \bigotimes_{k=1}^{n} O_{2k} ,
\end{equation}
where each $O_i$ denotes a linear operator on a single-qubit Hilbert space.

The orthonormality condition and the completeness relation for the basis $|\Phi_{2n}(\underline{\alpha\beta})\rangle$ give rise to two important properties of the $n$-fold tensor products of Pauli gates $T_n(\underline{\alpha\beta})$. First, the orthonormality relation between $T_n(\underline{\alpha\beta})$ and $T_n(\underline{\alpha^\prime\beta^\prime})$ with respect to the Hilbert--Schmidt inner product is
\begin{equation}
	\frac {1} {2^n} 	\operatorname{tr} \bigl(T_n^\dagger(\underline{\alpha^\prime\beta^\prime})   T_n(\underline{\alpha\beta}) \bigr)
	=\delta_{\underline{\alpha}^\prime\,\underline{\alpha}}
	\delta_{\underline{\beta}^\prime\,\underline{\beta}},
\end{equation}
with the convention for the Kronecker deltas 
\begin{equation}
	\delta_{\underline{\alpha}^\prime\,\underline{\alpha}}
	\equiv\prod_{k=1}^n\,\delta_{\alpha_k^\prime\alpha_k}, \qquad 
	\delta_{\underline{\beta}^\prime\,\underline{\beta}}
	\equiv \prod_{k=1}^n\,\delta_{\beta_k^\prime\beta_k},
\end{equation}
which shows that the set $\{T_n(\underline{\alpha\beta})\}$ forms an orthonormal basis for the $4^n$-dimensional Hilbert space. Second, the completeness relation yields
\begin{equation}
	\frac {1} {2^n} \sum_{\underline{\alpha},\underline{\beta}}
	T_n(\underline{\alpha\beta})
	|\underline{i}\rangle\langle \underline{j}|
	T_n^\dagger(\underline{\alpha\beta})
	=  \delta_{\underline{i}\,\underline{j}} 1\!\!1_2^{\otimes n},
\end{equation}
with the summation convention
\begin{equation}
	\sum_{\underline{\alpha},\underline{\beta}} \equiv  \sum_{\alpha_1,\dots,\alpha_n=0}^1 
	\sum_{\beta_1,\dots,\beta_n=0}^1, \quad 
	\delta_{\underline{i}\,\underline{j}}
	\equiv  \prod_{k=1}^n\,\delta_{i_k j_k}.
\end{equation}
These two properties, together with the unitarity of the twist operator $\tau_{2n}$, confirm that the basis $|\mathcal{B}_{2n}(\underline{\alpha\beta})\rangle$ is orthonormal.
In what follows, we focus on the generalized $2n$-qubit basis $|\mathcal{B}_{2n}(\underline{\alpha\beta})\rangle$, as it plays a key role in quantum teleportation of multiple qubits.

\subsection{Quantum circuit to generate  
	\texorpdfstring{\(|\mathcal{B}_{2n}(\underline{\alpha\beta})\rangle\)}{}}

\begin{figure}[!hbt]
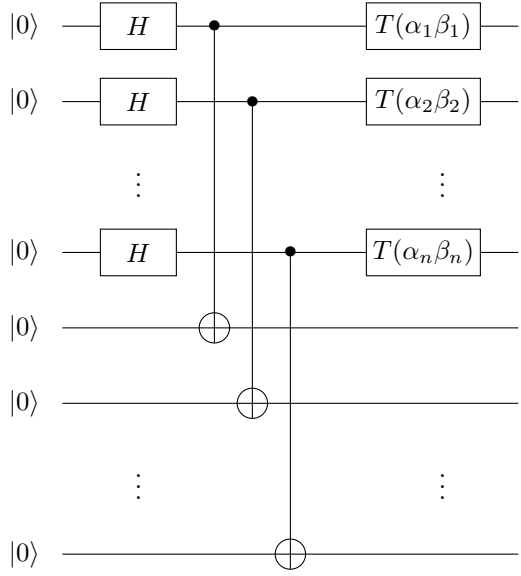

	\begin{equation*}
		\tikz[baseline]{
			\draw(0,8)--(0.5,8);
			\draw(0.5,8.309)rectangle(1.5,7.691);
			\draw(1.5,8)--(4,8);
			\draw(0,7)--(0.5,7);
			\draw(0.5,7.309)rectangle(1.5,6.691);
			\draw(1.5,7)--(4,7);
			\draw(1,6)coordinate(A) node {$\vdots$};
			\draw(5,6)coordinate(A) node {$\vdots$};
			\draw(0,5)--(0.5,5);
			\draw(0.5,5.309)rectangle(1.5,4.691);
			\draw(1.5,5)--(4,5);
			\draw(4,8.309)rectangle(5.5,7.691);
			\draw(4,7.309)rectangle(5.5,6.691);
			\draw(4,5.309)rectangle(5.5,4.691);
			\draw(5.5,8)--(6,8);
			\draw(5.5,7)--(6,7);
			\draw(5.5,5)--(6,5);
			\draw(0,4)--(6,4);
			\draw(0,3)--(6,3);
			\draw(1,2)coordinate(A) node {$\vdots$};
			\draw(5,2)coordinate(A) node {$\vdots$};
			\draw(0,1)--(6,1);
			\draw(2,8)coordinate(A) node {$\bullet$};
			\draw(2.5,7)coordinate(A) node {$\bullet$};
			\draw(3,5)coordinate(A) node {$\bullet$};
			\draw(2,8)--(2,3.8);
			\draw(2,4)circle (0.2);
			\draw(2.5,7)--(2.5,2.8);
			\draw(2.5,3)circle (0.2);
			\draw(3,5)--(3,0.8);
			\draw(3,1)circle (0.2);
			\draw(1,8)coordinate(A) node {$H$};
			\draw(1,7)coordinate(A) node {$H$};
			\draw(1,5)coordinate(A) node {$H$};
			\draw(4.75,8)coordinate(A) node {$T(\alpha_1\beta_1)$};
			\draw(4.75,7)coordinate(A) node {$T(\alpha_2\beta_2)$};
			\draw(4.75,5)coordinate(A) node {$T(\alpha_n\beta_n)$};
			\draw(-0.5,8)coordinate(A) node {$|0\rangle$};
			\draw(-0.5,7)coordinate(A) node {$|0\rangle$};
			\draw(-0.5,5)coordinate(A) node {$|0\rangle$};
			\draw(-0.5,4)coordinate(A) node {$|0\rangle$};
			\draw(-0.5,3)coordinate(A) node {$|0\rangle$};
			\draw(-0.5,1)coordinate(A) node {$|0\rangle$};
			
		}
	\end{equation*}
	\caption{Quantum circuit to create the generalized $2n$-qubit Bell state $|\mathcal{B}_{2n}(\underline{\alpha\beta})\rangle$.}
	\label{fig: QC_generalized-2n-bell-state}
\end{figure}

A quantum circuit for generalized $2n$-qubit Bell states $|\mathcal{B}_{2n}(\underline{\alpha\beta})\rangle$ is illustrated in Fig.~\ref{fig: QC_generalized-2n-bell-state}. First, using only Hadamard gates $H$ and CNOT gates, the quantum circuit for $|\mathcal{B}_{2n}\rangle$ is algebraically expressed as 
\begin{equation}
	|\mathcal{B}_{2n}\rangle=\bigotimes_{k=1}^n CNOT_{k,n+k}
	( H^{\otimes n}\otimes  1\!\! 1_2^{\otimes n}) |0\rangle^{\otimes 2n},
\end{equation}
where $CNOT_{k,n+k}$ has the $k$th qubit as the control and the $(n+k)$th qubit as the target, and the product of a sequence of CNOT gates is denoted compactly as
\begin{equation}
	\bigotimes_{k=1}^n CNOT_{k,n+k}
	=CNOT_{n,2n}\cdots CNOT_{1,n+1}.
\end{equation}
Second, tensor products of Pauli gates, $T_n(\underline{\alpha\beta})\otimes 1\!\! 1_2^{\otimes n}$, are applied to this state, thereby generating $|\mathcal{B}_{2n}(\underline{\alpha\beta})\rangle$ and completing the quantum circuit.

\subsection{The basis theorem for generalized \texorpdfstring{\(2n\)}{}-qubit Bell states}

\label{basis-theorem-multi-qubit}

Given \(2^n \times 2^n\) matrices \(M, L, N\) satisfying \(M=L N^\intercal\), the local action of \(M\) on the state \(|\mathcal{B}_{2n}\rangle\) has the form
\begin{equation}
	\begin{aligned}
		|\mathcal{M}_{2n}\rangle
		& \equiv  (M\otimes1\!\!1_2^{\otimes n}) |\mathcal{B}_{2n}\rangle\\
		&=(L\otimes N)  |\mathcal{B}_{2n}\rangle.
	\end{aligned}
\end{equation}
Provided \(M, L,\) and \(N\) are unitary matrices, the state \(|\mathcal{M}_{2n}\rangle\) has the Schmidt decomposition form of a bipartite pure state, and is therefore a maximally entangled state in the bipartite \(n\)-qubit Hilbert space. Analogously, denote \(|\mathcal{M}_{2n}^\intercal\rangle \equiv  (M^\intercal\otimes1\!\!1_2^{\otimes n}) |\mathcal{B}_{2n}\rangle\). 

With a \(2^n \times 2^n\) unitary matrix \(M\), the generalized \(2n\)-qubit Bell basis \(|\mathcal{B}_{2n}(\underline{\alpha\beta})\rangle\) can be naturally extended in several ways. The first extension has the form 
\begin{equation} 
	|M\mathcal{B}_{2n}(\underline{\alpha\beta})\rangle
	=(MT_n(\underline{\alpha\beta})\otimes1\!\!1_2^{\otimes n})|\mathcal{B}_{2n}\rangle,
\end{equation}
the second is given by
\begin{equation}
	|\mathcal{B} M_{2n}(\underline{\alpha\beta})\rangle
	=(T_n(\underline{\alpha\beta})M\otimes1\!\!1_2^{\otimes n})|\mathcal{B}_{2n}\rangle,
\end{equation}
and the third is
\begin{equation}
	|\mathcal{B} M^\intercal_{2n}(\underline{\alpha\beta})\rangle
	=(T_n(\underline{\alpha\beta})\otimes M)|\mathcal{B}_{2n}\rangle.
\end{equation}
Generally, \(M\) need not be an \(n\)-fold tensor product of single-qubit gates \(M_i\), i.e., \(M=M_1\otimes \cdots \otimes M_n\), with each \(M_i\) acting on the \(i\)-th qubit Hilbert space. 

As a crucial application of Theorem~\ref{basislemma}, we state the following corollary for generalized \(2n\)-qubit Bell states.  
\begin{corollary}
	Let \(M\) be a \(2^n \times 2^n\) matrix on the \(n\)-qubit Hilbert space. The set of states
	\(\{|M\mathcal{B}_{2n}(\underline{\alpha\beta})\rangle\}\) 
	(or \(\{\,|\mathcal{B} M_{2n}(\underline{\alpha\beta})\rangle\}\)) forms an orthonormal basis for the \(2n\)-qubit Hilbert space \(\mathscr{H}_{2^n}\otimes \mathscr{H}_{2^n}\) if and only if \(M\) is a unitary matrix.
\end{corollary}
The key step in the proof of the above corollary is to verify that the constraint equations derived from the orthogonality of the extended generalized \(2n\)-qubit Bell bases,
\begin{equation}
	\frac {1} {2^n} \operatorname{tr}\bigl(M M^\dagger T_n^\dagger(\underline{\alpha^\prime\beta^\prime})  
	T_n(\underline{\alpha\beta}) \bigr)
	=\delta_{\underline{\alpha}^\prime\,\underline{\alpha}}
	\delta_{\underline{\beta}^\prime\,\underline{\beta}},
\end{equation}
and  
\begin{equation}
	\frac {1} {2^n} 	\operatorname{tr}\bigl(M^\dagger M T_n(\underline{\alpha\beta}) T_n^\dagger(\underline{\alpha^\prime\beta^\prime}) \bigr)
	=\delta_{\underline{\alpha}^\prime\,\underline{\alpha}}
	\delta_{\underline{\beta}^\prime\,\underline{\beta}},
\end{equation}
imply the unitarity of \(M\).

First, one must verify that the set of all \(\pm T_n^\dagger(\underline{\alpha^\prime\beta^\prime}) T_n(\underline{\alpha\beta})\) forms the basis group defined in Subsec.~\ref{basisgroup}. Note that the Hermitian conjugate of \(T_n(\underline{\alpha^\prime\beta^\prime})\) remains in the set of all \(\pm T_n(\underline{\alpha\beta})\), since
\begin{equation}
	T_n^\dagger(\underline{\alpha^\prime\beta^\prime}) =(-1)^{\underline{\alpha^\prime}\cdot \underline{\beta^\prime}} T_n(\underline{\alpha^\prime\beta^\prime}),
\end{equation}
where the dot product between two \(n\)-bit strings \(\underline{\alpha^\prime}\) and \(\underline{\beta^\prime}\) is defined as \(\underline{\alpha^\prime}\cdot \underline{\beta^\prime} \equiv
\alpha_1^\prime\beta_1^\prime\oplus\cdots\oplus \alpha_n^\prime\beta_n^\prime\) with bit multiplication \(\alpha_i^\prime\beta_i^\prime\) and bit addition \(\oplus\). Moreover, the product
\(T_n^\dagger(\underline{\alpha^\prime\beta^\prime}) T_n(\underline{\alpha\beta})\) is itself an element of this set:
\begin{equation}
	T_n^\dagger(\underline{\alpha^\prime\beta^\prime})  T_n(\underline{\alpha\beta})
	=(-1)^{\underline{\alpha^\prime}\cdot \underline{\beta^\prime}}
	(-1)^{\underline{\alpha}\cdot \underline{\beta^\prime}} T_n(\underline{\alpha^{\prime\prime}\beta^{\prime\prime }}),
\end{equation}
with \(\alpha_i^{\prime\prime}=\alpha_i\oplus \alpha_i^\prime\) for the \(i\)th bit of \(\underline{\alpha^{\prime\prime}}\) and
\(\beta_i^{\prime\prime }=\beta_i\oplus \beta_i^{\prime }\) for the \(i\)th bit of
\(\underline{\beta^{\prime\prime }}\). 
Hence the set of all \(\pm T_n(\underline{\alpha\beta})\) is closed under conjugate transpose and multiplication. As a result, verifying the corollary reduces to proving the following lemma.

\begin{lemma}	\label{lemmainductiveproof}
	The constraint condition 
	\begin{equation}
		\frac {1} {2^n} 	\operatorname{tr}\bigl(I(n) \,\, T_n(\underline{\alpha\beta}) \bigr)
		=\delta_{\underline{\alpha}\,\underline{0}}
		\delta_{\underline{\beta}\,\underline{0}},
	\end{equation}
	with the \(n\)-bit string \(\underline{0}=0\cdots 0\), implies that the matrix \(I(n)\) is the identity \(1\!\!1_{2^n}\).
\end{lemma}

A direct proof of the lemma is as follows. Since the set of \(\pm T_n(\underline{\alpha\beta})\) is an orthonormal basis for the \(4^n\)-dimensional Hilbert space equipped with the Hilbert--Schmidt inner product, the matrix \(I(n)\) must be the identity \(1\!\!1_2^{\otimes n}\). An alternative algebraic proof by induction is provided in Appendix~\ref{lianmingproof}.

In addition to the above approach based on the orthonormality condition, the unitarity of \(M\) can also be derived from the completeness relation for the orthonormal bases. The completeness relation for \(|M\mathcal{B}_{2n}(\underline{\alpha\beta})\rangle\) directly implies the unitarity of \(M\). Furthermore, using the reduced completeness relation for \(|\mathcal{B} M_{2n}(\underline{\alpha\beta})\rangle\), given by
\begin{equation}
	\frac {1} {2^n} \sum_{\underline{\alpha},\underline{\beta}}
	T_n(\underline{\alpha\beta}) M
	|\underline{i}\rangle\langle \underline{j}| M^\dagger
	T_n^\dagger(\underline{\alpha\beta})
	=  (M^\dagger M)_{\underline{j}\,\underline{i}} 1\!\!1_2^{\otimes n},
\end{equation}
one can readily verify the corollary. 

Note that the \(2^n \times 2^n\) matrix \(M\) acting on the \(n\)-qubit state \(|\underline{i}\rangle\) has the form
\begin{equation}
	M|\underline{i}\rangle=\sum_{\underline{k}=0}^{2^n-1} |\underline{k}\rangle 
	M_{\underline{k}\,\underline{i}} ,
\end{equation}
where \(M_{\underline{k}\,\underline{i}}\) denotes the matrix element labeled by the \(n\)-bit strings \(\underline{k}\) and \(\underline{i}\):
\begin{equation}
	M_{\underline{k}\,\underline{i}}=M_{k_1k_2\cdots k_n, i_1 i_2 \cdots i_n},
\end{equation}
and it is often a linear combination of \(n\)-fold tensor products of single-qubit gates.

\subsection{Observables for generalized \texorpdfstring{$2n$}{}-qubit Bell states}

A complete set of observables specifying the generalized $2n$-qubit Bell states $|\mathcal{B}_{2n}(\underline{\alpha\beta})\rangle$ as eigenvectors includes the phase-bit observables $X_k\otimes X_{n+k}$ and the parity-bit observables $Z_k\otimes Z_{n+k}$, for $k=1,\ldots, n$. The subscript of a Pauli gate indicates the qubit on which it acts; for example, $X_k$ and $Z_k$ are applied to the $k$th qubit. The observables, eigenvectors, and eigenvalues are given by
\begin{equation}\label{Xi and Zi}
	\begin{aligned}
		&(X_k\otimes X_{n+k}) |\mathcal{B}_{2n}(\underline{\alpha\beta})\rangle 
		=(-1)^{\alpha_k} |\mathcal{B}_{2n}(\underline{\alpha\beta})\rangle,\\
		&(Z_k\otimes Z_{n+k})|\mathcal{B}_{2n}(\underline{\alpha\beta})\rangle =(-1)^{\beta_k}|\mathcal{B}_{2n}(\underline{\alpha\beta})\rangle,
	\end{aligned}
\end{equation}
and the eigenvalues encode the $n$-bit strings $\underline{\alpha}$ and $\underline{\beta}$ through their respective exponents.

Following the construction of a complete set of observables for the eigenvectors $|M\Omega(\alpha\beta)\rangle$ (or $|\Omega M(\alpha\beta)\rangle$) detailed in (\ref{M_OX+k}) and (\ref{M_OZ+k}), we apply the unitary transformation $M\otimes 1\!\!1_2^{\otimes n}$ (or $1\!\!1_2^{\otimes n} \otimes M^{\intercal}$) to the above observables. This yields the complete set of observables that define $|M\mathcal{B}_{2n}(\underline{\alpha\beta})\rangle$ (or $|\mathcal{B} M_{2n}(\underline{\alpha\beta})\rangle$) as eigenvectors.

\subsection{Calculation of generalized concurrence via generalized Bell states}

The generalized concurrence \cite{Hill1997, Coffman2000, Wong2001} is a well-established entanglement measure for pure states. Rigolin \cite{Rigolin_2005} provided a brief calculation of the generalized concurrence for a four-qubit pure state expressed as a linear combination of four-qubit Bell states. Here, we compute the generalized concurrence of a \(2n\)-qubit pure state \(|\underline{\psi}_{2n}\rangle\) expanded as a linear combination of generalized \(2n\)-qubit Bell states \(|\mathcal{B}_{2n}(\underline{\alpha\beta})\rangle\).

The generalized concurrence of \(|\underline{\psi}_{2n}\rangle\) is defined by
\begin{equation}
	\mathcal{C}(|\underline{\psi}_{2n}\rangle) \equiv 
	|\langle\widetilde{\underline{\psi}}_{2n}|\underline{\psi}_{2n}\rangle|,
\end{equation}
with the spin-flip state
\(|\widetilde{\underline{\psi}}_{2n}\rangle =(-1)^n (ZX)^{\otimes 2n}|\underline{\psi}_{2n}^*\rangle\). 
The state \(|\underline{\psi}_{2n}\rangle\) has the form    
\begin{equation}
	|\underline{\psi}_{2n}\rangle=\sum_{\underline{\alpha},\underline{\beta}}
	d_{2n}(\underline{\alpha\beta}) |\mathcal{B}_{2n}(\underline{\alpha\beta})\rangle,
\end{equation}
with the normalization condition \(\sum_{\underline{\alpha},\underline{\beta}} |d_{2n}(\underline{\alpha\beta})|^2=1\). The spin-flip state \(|\widetilde{\underline{\psi}}_{2n}\rangle\) is then
\begin{equation}
	|\widetilde{\underline{\psi}}_{2n}\rangle
	=(-1)^n \sum_{\underline{\alpha},\underline{\beta}} (-1)^{\sum_{k=1}^n(\alpha_k \oplus\beta_k)} d^*_{2n}(\underline{\alpha\beta})|\mathcal{B}_{2n}(\underline{\alpha\beta})\rangle,
\end{equation}
where we have used the identity
\begin{equation}
	(ZX)^{\otimes n} T_n(\underline{\alpha\beta}) (XZ)^{\otimes n} 
	=(-1)^{\sum_{k=1}^n(\alpha_k \oplus \beta_k)} T_n(\underline{\alpha\beta}).
\end{equation}
Hence, the generalized concurrence of \(|\underline{\psi}_{2n}\rangle\) is given by the compact formula
\begin{equation}
	\label{concurrence formula}
	\mathcal{C}(|\underline{\psi}_{2n}\rangle)=
	\left|\sum_{\underline{\alpha},\underline{\beta}} (-1)^{\sum_{k=1}^n(\alpha_k \oplus \beta_k)} 
	\bigl(d_{2n}(\underline{\alpha\beta})\bigr)^2\right|.
\end{equation}

As a first example, a generalized \(2n\)-qubit Bell state \(|\mathcal{B}_{2n}(\underline{\alpha\beta})\rangle\) has concurrence
\(\mathcal{C}(|\mathcal{B}_{2n}(\underline{\alpha\beta})\rangle)=1\), as expected, confirming that it is a maximally entangled bipartite pure state.

Second, we compute the concurrence of a \(2n\)-qubit product basis state 
\(|\underline{j} \underline{l} \rangle \equiv  |j_1\cdots j_n l_1\cdots l_n \rangle\), although its concurrence is known to be zero. Using the identity
\begin{equation}
	T(\alpha_k \beta_k) |i_k\rangle = (-1)^{\alpha_k (i_k \oplus \beta_k)} |i_k\rangle,
\end{equation}
we rewrite \(|\underline{j} \underline{l} \rangle\) as a linear combination of \(|\mathcal{B}_{2n}(\underline{\alpha\beta})\rangle\):
\begin{equation}
	|\underline{j} \underline{l}\rangle
	=\frac{1}{\sqrt{2^n}} \sum_{\underline{\alpha},\underline{\beta}} (-1)^{\sum_{k=1}^n \alpha_k j_k} \delta_{\underline{\beta}, \underline{l}\oplus \underline{j}}  |\mathcal{B}_{2n}(\underline{\alpha\beta})\rangle,
\end{equation}
with \(\delta_{\underline{\beta}, \underline{l}\oplus \underline{j}} = \delta_{\beta_1, l_1\oplus j_1} \cdots \delta_{\beta_n, l_n\oplus j_n}\). Thus \(\bigl(d_{2n}(\underline{\alpha\beta})\bigr)^2\) in \eqref{concurrence formula} becomes  
\begin{equation}
	\bigl(d_{2n}(\underline{\alpha\beta})\bigr)^2=\frac{1}{2^n} \delta_{\underline{\beta}, \underline{l}\oplus \underline{j}},
\end{equation}
and the generalized concurrence of \(|\underline{j} \underline{l} \rangle\) is indeed zero:
\begin{equation}
	\mathcal{C}(|\underline{j} \underline{l} \rangle)=
	\left|\sum_{\underline{\alpha}} (-1)^{\sum_{k=1}^n \alpha_k }\right|=0.
\end{equation}

Third, we calculate the generalized concurrence of the \(2n\)-qubit GHZ states
\begin{equation}
	|GHZ_{2n}(\pm)\rangle=\frac{1}{\sqrt{2}}\bigl(|j_1\cdots j_n l_1\cdots l_n \rangle 
	\pm |\overline{j}_1\cdots \overline{j}_n \overline{l}_1\cdots \overline{l}_n \rangle\bigr),
\end{equation}
with \(\overline{j}_k=1\oplus j_k\) and \(\overline{l}_k=1 \oplus l_k\), satisfying \(j_k \oplus l_k =\overline{j}_k \oplus \overline{l}_k\). In terms of the generalized Bell states \(|\mathcal{B}_{2n}(\underline{\alpha\beta})\rangle\), the GHZ states are expressed as 
\begin{equation}
	|GHZ_{2n}(\pm)\rangle=\sum_{\underline{\alpha},\underline{\beta}} d^\pm_{2n}(\underline{\alpha\beta})  |\mathcal{B}_{2n}(\underline{\alpha\beta})\rangle,
\end{equation}
with amplitudes
\begin{equation}
	d^\pm_{2n}(\underline{\alpha\beta})=\frac{1}{\sqrt{2^{n+1}}}  
	\Bigl((-1)^{\sum_{k=1}^n \alpha_k j_k} \pm (-1)^{\sum_{k=1}^n \alpha_k \overline{j}_k} \Bigr)
	\delta_{\underline{\beta}, \underline{l}\oplus \underline{j}}. 
\end{equation}
After some algebra, using
\begin{equation}
	\bigl(d^\pm_{2n}(\underline{\alpha\beta})\bigr)^2=\frac{1}{2^{n}}  
	\bigl(1\pm (-1)^{\sum_{k=1}^n \alpha_k} \bigr)
	\delta_{\underline{\beta}, \underline{l}\oplus \underline{j}},
\end{equation}
we obtain the concurrence of the GHZ states:
\begin{equation}
	\mathcal{C}(|GHZ_{2n}(\pm)\rangle)=\frac{1}{\sqrt{2^{n}}}  
	\left|\sum_{\underline{\alpha}} (-1)^{\sum_{k=1}^n \alpha_k }\bigl(1 \pm (-1)^{\sum_{k=1}^n \alpha_k }\bigr)\right|=1,
\end{equation}
which confirms that the GHZ states are indeed maximally entangled.

\section{Diagrammatic representations of generalized Bell states}

This section presents graphical descriptions of generalized Bell states within the extended Temperley--Lieb diagrammatic approach \cite{Zhang2006,Zhang2009}. In addition to a brief introduction to the Temperley--Lieb algebra and its representation, various diagrams are drawn to illustrate algebraic properties of generalized two-qudit Bell states, including the orthonormality condition and the completeness relation. Using schematic representations of SWAP gates and two-qubit Bell states, we then construct an illustration of generalized $2n$-qubit Bell states.

\subsection{The Temperley--Lieb algebra and the basis theorem}

The Temperley--Lieb algebra plays an essential role in both statistical mechanics and low-dimensional topology, as detailed in \cite{Kauffman2012}. It has been well demonstrated \cite{Zhang2006,Zhang2009} that the extended Temperley--Lieb diagrammatic approach is capable of describing quantum information protocols involving generalized Bell states and local quantum gates, such as the quantum teleportation discussed in this paper. Here we briefly sketch the Temperley--Lieb algebra and its representation.      

The Temperley--Lieb algebra $TL_n(\lambda)$ with loop parameter $\lambda$ is generated by a set of idempotents $e_1,e_2,\dots,e_{n-1}$ satisfying $e_i^2=e_i$ and 
\begin{equation}\label{87}
	e_i e_{i\pm 1} e_i=\lambda^{-2}e_i,
\end{equation}
and  $e_j e_k=e_k e_j$ for $|j-k|> 1$ with $j,k=1,2,\dots,n-1$. As in previous studies \cite{Zhang2006,Zhang2009}, an interesting representation of the Temperley--Lieb algebra $TL_n(d)$ can be constructed via generalized two-qudit Bell states $|\Omega(a)\rangle$:
\begin{equation}\label{generator}
	e_i=1\!\!1_d^{\otimes (i-1)}\otimes |\Omega(a)\rangle\langle\Omega(a)|\otimes 1\!\!1_d^{\otimes (n-i-1)}.
\end{equation}
Similarly, utilizing $|M\Omega(a)\rangle$ or $|\Omega M(a)\rangle$ with a unitary matrix $M$, one can formulate a representation of the Temperley--Lieb algebra $TL_n(d)$. 

To establish the connection between the Temperley--Lieb algebra and the basis theorem, we state the following theorem.
\begin{theorem}
	A representation of the Temperley--Lieb algebra \(TL_n(d)\) can be constructed via generalized two-qudit Bell states \(|M\Omega(a)\rangle\) or \(|\Omega M(a)\rangle\) if and only if the \(d \times d\) matrix \(M\) is unitary.
\end{theorem}
For details on the Temperley--Lieb algebra and its representation theory, interested readers are referred to Kauffman's book \cite{Kauffman2012}.

\subsection{The extended Temperley--Lieb diagrammatic approach}

A brief description of the extended Temperley--Lieb diagrammatic approach \cite{Zhang2006,Zhang2009} is provided for handling diagrammatic teleportation in the subsequent sections. In addition to cups, caps, straight lines, inclined lines, triangular symbols, black dots, etc., various graphical elements are needed to construct a complete picture of a quantum information protocol using generalized Bell states. In particular, algebraic properties of generalized two-qudit Bell states are illustrated through different combinations of cups and caps.

First, one draws a picture from left to right and from top to bottom. 

\begin{figure}[!hbt]
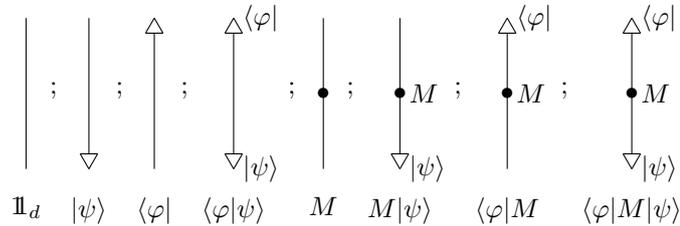

	\begin{equation*}
		%\begin{equation}\label{Rule 1}
		\tikz[baseline]{
			\draw(0,-1)--(0,1);
			\draw(0,-1.25)coordinate(A)node[below]{$1\!\!1_d$};
		}
		;
		\tikz[baseline]{
			\draw(0,-0.8)--(0,1);
			\draw(-0.115,-0.8)--(0.115,-0.8);
			\draw(-0.115,-0.8)--(0,-1);
			\draw(0.115,-0.8)--(0,-1);
			%	\draw(0,-.6)coordinate(A)node[below]{$\triangledown$};
			\draw(0,-1.25)coordinate(A)node[below]{$|\psi\rangle$};
		}
		;
		\tikz[baseline]{
			\draw(0,-1)--(0,0.8);
			\draw(-0.115,0.8)--(0.115,0.8);
			\draw(-0.115,0.8)--(0,1);
			\draw(0.115,0.8)--(0,1);
			\draw(0,-1.25)coordinate(A)node[below]{$\langle\varphi|$};
		}
		;
		\tikz[baseline]{
			\draw(0,-0.8)--(0,0.8);
			\draw(-0.115,-0.8)--(0.115,-0.8);
			\draw(-0.115,-0.8)--(0,-1);
			\draw(0.115,-0.8)--(0,-1);
			\draw(-0.115,0.8)--(0.115,0.8);
			\draw(-0.115,0.8)--(0,1);
			\draw(0.115,0.8)--(0,1);
			\draw(0,-1.25)coordinate(A)node[below]{$\langle\varphi|\psi\rangle$};
			\draw(0,1)coordinate(A)node[right]{$\langle\varphi|$};
			\draw(0,-1)coordinate(A)node[right]{$|\psi\rangle$};
		}
		;
		\tikz[baseline]{
			\draw(0,-1)--(0,1);
			\draw(0,-1.25)coordinate(A)node[below]{$M$};
			\draw(0,0)coordinate(A)node{$\bullet$};
		}
		;
		\tikz[baseline]{
			\draw(0,-0.8)--(0,1);
			\draw(-0.115,-0.8)--(0.115,-0.8);
			\draw(-0.115,-0.8)--(0,-1);
			\draw(0.115,-0.8)--(0,-1);
			\draw(0,-1.25)coordinate(A)node[below]{$M|\psi\rangle$};
			\draw(0,0)coordinate(A)node{$\bullet$};
			\draw(0,0)coordinate(A)node[right]{$M$};
			\draw(0,-1)coordinate(A)node[right]{$|\psi\rangle$};
		}
		;
		\tikz[baseline]{
			\draw(0,-1)--(0,0.8);
			\draw(-0.115,0.8)--(0.115,0.8);
			\draw(-0.115,0.8)--(0,1);
			\draw(0.115,0.8)--(0,1);
			\draw(0,-1.25)coordinate(A)node[below]{$\langle\varphi|M$};
			\draw(0,0)coordinate(A)node{$\bullet$};
			\draw(0,0)coordinate(A)node[right]{$M$};
			\draw(0,1)coordinate(A)node[right]{$\langle\varphi|$};
		}
		;
		\tikz[baseline]{
			\draw(0,-0.8)--(0,0.8);
			\draw(-0.115,-0.8)--(0.115,-0.8);
			\draw(-0.115,-0.8)--(0,-1);
			\draw(0.115,-0.8)--(0,-1);
			\draw(-0.115,0.8)--(0.115,0.8);
			\draw(-0.115,0.8)--(0,1);
			\draw(0.115,0.8)--(0,1);
			\draw(0,0)coordinate(A)node{$\bullet$};
			\draw(0,0)coordinate(A)node[right]{$M$};
			\draw(0,-1.25)coordinate(A)node[below]{$\langle\varphi|M|\psi\rangle$};
			\draw(0,1)coordinate(A)node[right]{$\langle\varphi|$};
			\draw(0,-1)coordinate(A)node[right]{$|\psi\rangle$};
		}
	\end{equation*}
	\caption{Rules for drawing operators, state vectors, and inner products.}
	\label{fig: Rule 1}
\end{figure}

See Fig.~\ref{fig: Rule 1}. A straight line represents the identity $1\!\!1_d$. A lower triangle $\triangledown$ at the endpoint of a ray denotes a state $|\psi\rangle$ (or $|\psi_d\rangle$), an upper triangle $\vartriangle$ at the endpoint of a ray denotes a state $\langle\varphi|$ (or $\langle\varphi_d|$), and the combination of an upper triangle and a lower triangle at the endpoints of a line segment indicates an inner product $\langle\varphi|\psi\rangle$. A black dot $\bullet$ placed in the middle of a straight line marks the local operator $M$, and the combinations of the black dot with triangular symbols describe $M|\psi\rangle$, $\langle\varphi|M$, and $\langle\varphi|M|\psi\rangle$, respectively.   

\begin{figure}[!hbt]
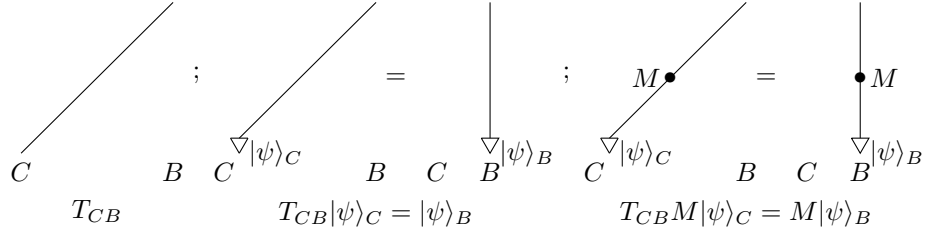

	\begin{equation*}
		%\begin{equation}	\label{Rule 2}
		\tikz[baseline]{
			\draw(-1,-1)--(1,1);
			\draw(-1,-1)coordinate(A)node[below]{$C$};
			\draw(1,-1)coordinate(A)node[below]{$B$};
			\draw(0,-1.5)coordinate(A)node[below]{${T}_{CB}$};
		}
		;
		\tikz[baseline]{
			\draw(-0.8,-0.8)--(1,1);		
			\draw(-0.115-0.8,-0.8)--(0.115-0.8,-0.8);
			\draw(-0.115-0.8,-0.8)--(0-0.8,-1);
			\draw(0.115-0.8,-0.8)--(0-0.8,-1);
			
			\draw(-1,-1)coordinate(A)node[below]{$C$};
			\draw(1,-1)coordinate(A)node[below]{$B$};
			\draw(1,-1.5)coordinate(A)node[below]{${T}_{CB}|\psi\rangle_C=|\psi\rangle_B$};
			\draw(1,0)coordinate(A)node[right]{$=$};
			\draw(1.8,-1)coordinate(A)node[below]{$C$};
			\draw(2.5,-1)coordinate(A)node[below]{$B$};
			\draw(2.5,-0.8)--(2.5,1);
			\draw(2.385,-0.8)--(2.615,-0.8)--(2.5,-1)--(2.385,-0.8);
			\draw(-0.8,-1)coordinate(A)node[right]{$|\psi\rangle_C$};
			\draw(2.5,-1)coordinate(A)node[right]{$|\psi\rangle_B$};
		}
		;
		\tikz[baseline]{
			\draw(-0.8,-0.8)--(1,1);
			\draw(-0.115-0.8,-0.8)--(0.115-0.8,-0.8);
			\draw(-0.115-0.8,-0.8)--(0-0.8,-1);
			\draw(0.115-0.8,-0.8)--(0-0.8,-1);
			%	\draw(-0.742,-0.858)--(-0.858,-0.742);
			%	\draw(-0.742,-0.858)--(-1,-1);
			%	\draw(-0.858,-0.742)--(-1,-1);
			\draw(-1,-1)coordinate(A)node[below]{$C$};
			\draw(1,-1)coordinate(A)node[below]{$B$};
			\draw(1,-1.5)coordinate(A)node[below]{${T}_{CB}M|\psi\rangle_C=M|\psi\rangle_B$};
			\draw(1,0)coordinate(A)node[right]{$=$};
			\draw(1.8,-1)coordinate(A)node[below]{$C$};
			\draw(2.5,-1)coordinate(A)node[below]{$B$};
			\draw(2.5,-0.8)--(2.5,1);
			\draw(2.385,-0.8)--(2.615,-0.8)--(2.5,-1)--(2.385,-0.8);
			\draw(-0.8,-1)coordinate(A)node[right]{$|\psi\rangle_C$};
			\draw(2.5,-1)coordinate(A)node[right]{$|\psi\rangle_B$};
			\draw(0,0)coordinate(A)node{$\bullet$};
			\draw(0,0)coordinate(A)node[left]{$M$};
			\draw(2.5,0)coordinate(A)node{$\bullet$};
			\draw(2.5,0)coordinate(A)node[right]{$M$};
		}
		%\end{equation}
	\end{equation*}
	\caption{Rules for drawing the transfer operator.}
	\label{fig: Rule 2}
\end{figure}

See Fig.~\ref{fig: Rule 2}. An inclined line from one state space $\mathscr{H}_{C}$ to another state space $\mathscr{H}_{B}$ represents the transfer operator $T_{CB}$ in \eqref{transfer operator1}, which maps the state vector $|\psi\rangle_{C}$ to the state vector $|\psi\rangle_{B}$, and similarly $M|\psi\rangle_{C}$ to $M|\psi\rangle_{B}$.

\begin{figure}[!hbt]
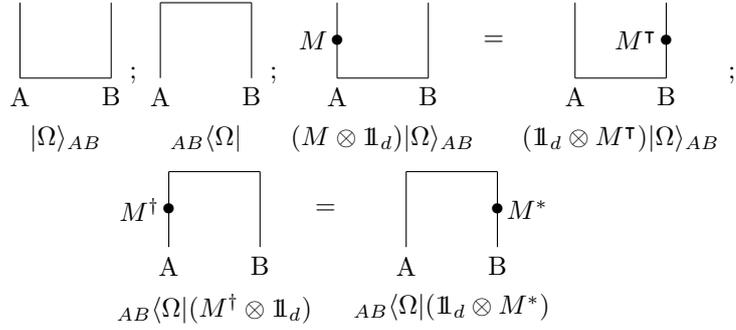

	\begin{equation*}
		%\begin{equation*}	%\label{Rule 3}
		\begin{aligned}
			&\tikz[baseline]{
				\draw(-0.6,1)--(-0.6,0);
				\draw(-0.6,0)--(0.6,0);
				\draw(0.6,0)--(0.6,1);
				\draw(-0.6,0)coordinate(A)node[below]{A};
				\draw(0.6,0)coordinate(A)node[below]{B};
				\draw(0,-0.5)coordinate(A)node[below]{$|\Omega\rangle_{AB}$};}
			;
			\tikz[baseline]{
				\draw(-0.6,1)--(-0.6,0);
				\draw(-0.6,1)--(0.6,1);
				\draw(0.6,0)--(0.6,1);
				\draw(-0.6,0)coordinate(A)node[below]{A};
				\draw(0.6,0)coordinate(A)node[below]{B};
				\draw(0,-0.5)coordinate(A)node[below]{$_{AB}\langle\Omega|$};}
			;
			\tikz[baseline]{
				\draw(-0.6,1)--(-0.6,0);
				\draw(-0.6,0)--(0.6,0);
				\draw(0.6,0)--(0.6,1);
				\draw(-0.6,0)coordinate(A)node[below]{A};
				\draw(0.6,0)coordinate(A)node[below]{B};
				\draw(0,-0.5)coordinate(A)node[below]{$({M}\otimes1\!\!1_d)|\Omega\rangle_{AB}$};
				\draw(-0.6,0.5)coordinate(A)node{$\bullet$};
				\draw(-0.6,0.5)coordinate(A)node[left]{${M}$};
				\draw(1.2,0.5)coordinate(A)node[right]{$=$};
			}
			\tikz[baseline]{
				\draw(-0.6,1)--(-0.6,0);
				\draw(-0.6,0)--(0.6,0);
				\draw(0.6,0)--(0.6,1);
				\draw(-0.6,0)coordinate(A)node[below]{A};
				\draw(0.6,0)coordinate(A)node[below]{B};
				\draw(0,-0.5)coordinate(A)node[below]{$(1\!\!1_d\otimes M^\intercal)|\Omega\rangle_{AB}$};
				\draw(0.6,0.5)coordinate(A)node{$\bullet$};
				\draw(0.6,0.5)coordinate(A)node[left]{${M}^\intercal$};
			}
			;\\
			&\qquad\qquad\tikz[baseline]{
				\draw(-0.6,1)--(-0.6,0);
				\draw(-0.6,1)--(0.6,1);
				\draw(0.6,0)--(0.6,1);
				\draw(-0.6,0)coordinate(A)node[below]{A};
				\draw(0.6,0)coordinate(A)node[below]{B};
				\draw(0,-0.5)coordinate(A)node[below]{$_{AB}\langle\Omega|({M}^\dagger\otimes1\!\!1_d)$};
				\draw(-0.6,0.5)coordinate(A)node{$\bullet$};
				\draw(-0.6,0.5)coordinate(A)node[left]{${M}^\dagger$};
				\draw(1.2,0.5)coordinate(A)node[right]{$=$};
			}
			\tikz[baseline]{
				\draw(-0.6,1)--(-0.6,0);
				\draw(-0.6,1)--(0.6,1);
				\draw(0.6,0)--(0.6,1);
				\draw(-0.6,0)coordinate(A)node[below]{A};
				\draw(0.6,0)coordinate(A)node[below]{B};
				\draw(0,-0.5)coordinate(A)node[below]{$_{AB}\langle\Omega|(1\!\!1_d\otimes{M}^*)$};
				\draw(0.6,0.5)coordinate(A)node{$\bullet$};
				\draw(0.6,0.5)coordinate(A)node[right]{${M}^*$};
			}
		\end{aligned}
	\end{equation*}
	\caption{Rules for drawing generalized two-qudit Bell states.}
	\label{fig: Rule 3}
\end{figure}

See Fig.~\ref{fig: Rule 3}.
A cup state in the two-qudit Hilbert space $\mathscr{H}_{A}\otimes \mathscr{H}_{B}$ represents a generalized two-qudit Bell state $|\Omega\rangle_{AB}$, and a cap state represents its Hermitian conjugate $_{AB}\langle\Omega|$. A local operator $M$ on $\mathscr{H}_{A}$ moves along the cup and becomes its transpose $M^\intercal$ on $\mathscr{H}_{B}$. Similarly, its Hermitian conjugate $M^\dagger$ on $\mathscr{H}_{A}$ moves along the cap and becomes its complex conjugate $M^*$ on $\mathscr{H}_{B}$.

\begin{figure}[!hbt]
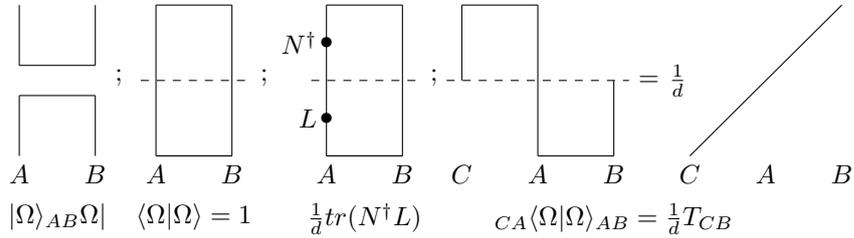

	\begin{equation*} 
		%	\begin{equation} \label{Rule 4}
			\tikz[baseline]{
				\draw(-0.5,1.0)--(-0.5,0.2);
				\draw(0.5,1.0)--(0.5,0.2);
				\draw(-0.5,0.2)--(0.5,0.2);
				\draw(-0.5,-1.)--(-0.5,-0.2);
				\draw(0.5,-1.)--(0.5,-0.2);
				\draw(-0.5,-0.2)--(0.5,-0.2);
				\draw(-0.5,-1.)coordinate(A)node[below]{$A$};
				\draw(0.5,-1.)coordinate(A)node[below]{$B$};
				\draw(0,-1.5)coordinate(A)node[below]{$|\Omega\rangle_{AB}\langle\Omega|$};
			}
			;
			\tikz[baseline]{
				\draw(-0.5,1)--(-0.5,-1);
				\draw(-0.5,-1)--(0.5,-1);
				\draw(0.5,-1)--(0.5,1);
				\draw(0.5,1)--(-0.5,1);
				\draw(-0.5,-1)--(0.5,-1);
				\draw[dashed](-0.7,0)--(0.7,0);
				\draw(-0.5,-1)coordinate(A)node[below]{$A$};
				\draw(0.5,-1)coordinate(A)node[below]{$B$};
				\draw(0,-1.5)coordinate(A)node[below]{$\langle\Omega|\Omega\rangle=1$};
			}
			;
			\tikz[baseline]{
				\draw(-0.5,1)--(-0.5,-1);
				\draw(-0.5,-1)--(0.5,-1);
				\draw(0.5,-1)--(0.5,1);
				\draw(0.5,1)--(-0.5,1);
				\draw(-0.5,-1)--(0.5,-1);
				\draw[dashed](-0.7,0)--(0.7,0);
				\draw(-0.5,-1)coordinate(A)node[below]{$A$};
				\draw(0.5,-1)coordinate(A)node[below]{$B$};
				\draw(0,-1.5)coordinate(A)node[below]{$\frac{1}{d}\operatorname{tr}({N}^\dagger{L})$};
				\draw(-0.5,0.5)coordinate(A)node{$\bullet$};
				\draw(-0.5,0.5)coordinate(A)node[left]{${N}^\dagger$};
				\draw(-0.5,-0.5)coordinate(A)node{$\bullet$};
				\draw(-0.5,-0.5)coordinate(A)node[left]{${L}$};
			}
			;
			\tikz[baseline]{
				\draw(-0.5,0)--(-0.5,1);
				\draw(-0.5,1)--(0.5,1);
				\draw(0.5,1)--(0.5,-1);
				\draw(1.5,-1)--(1.5,0);
				\draw(0.5,-1)--(1.5,-1);
				\draw[dashed](-0.7,0)--(1.7,0);
				\draw(-0.5,-1)coordinate(A)node[below]{$C$};
				\draw(0.5,-1)coordinate(A)node[below]{$A$};
				\draw(1.5,-1)coordinate(A)node[below]{$B$};
				\draw(1.7,0)coordinate(A)node[right]{$=$};
				\draw(2.1,0)coordinate(A)node[right]{$\frac{1}{d}$};
				%	\draw(2.5,1)--(4.5,-1);
				\draw(2.5,-1)--(4.5,1);
				\draw(2.5,-1)coordinate(A)node[below]{$C$};
				\draw(3.5,-1)coordinate(A)node[below]{$A$};
				\draw(4.5,-1)coordinate(A)node[below]{$B$};
				\draw(1.5,-1.5)coordinate(A)node[below]{$_{CA}\langle\Omega|\Omega\rangle_{AB}=\frac{1}{d}{T}_{CB}$};
			}
			%	\end{equation}
	\end{equation*}
	\caption{Various combinations of cup and cap states.}
	\label{fig: Rule 4}
\end{figure}

See Fig.~\ref{fig: Rule 4}.
An upper cup with a lower cap on the space $\mathscr{H}_{A}\otimes \mathscr{H}_{B}$ represents $|\Omega\rangle_{AB}\langle\Omega|$. An upper cap with a lower cup forms a loop, describing an inner product between the two states or the trace of a matrix product. The disappearance of a cup and a cap yields the factor $1/d$, and the trace of the identity matrix gives the factor $d$, so that a loop without any dots simply indicates that the state $|\Omega\rangle_{AB}$ is normalized.

As shown above, an upper cap on the left in the space $\mathscr{H}_{C}\otimes \mathscr{H}_{A}$ together with a lower cup on the right in the space $\mathscr{H}_{A}\otimes \mathscr{H}_{B}$ can be diagrammatically deformed into an inclined line representing the transfer operator $\frac{1}{d} T_{CB}$ from \eqref{transfer operator2}. This graphical operation closely resembles those used in low-dimensional topology \cite{Kauffman2012} and plays an essential role in the following schematic description of quantum teleportation.

\subsection{Diagrammatic orthonormality condition and completeness relation}

For generalized two-qudit Bell states, consider the outer product $|\Omega(b)\rangle_{AB}\langle\Omega(a)|$ shown in Fig.~\ref{fig: outer product}.
\begin{figure}[!hbt]
	\begin{equation*} 
		%\begin{equation*}\label{diag Bell basis}
		\tikz[baseline]{
			\draw(-0.5,1.2)--(-0.5,0.2);
			\draw(0.5,1.2)--(0.5,0.2);
			\draw(-0.5,0.2)--(0.5,0.2);
			\draw(-0.5,-1.2)--(-0.5,-0.2);
			\draw(0.5,-1.2)--(0.5,-0.2);
			\draw(-0.5,-0.2)--(0.5,-0.2);
			\draw(-0.5,-1.2)coordinate(A)node[below]{$A$};
			\draw(0.5,-1.2)coordinate(A)node[below]{$B$};
			\draw(-2.5,0)coordinate(A) node {$|\Omega(b)\rangle_{AB}\langle\Omega(a)|=$};
			\draw(-0.5,0.7)coordinate(A) node {$\bullet$};
			\draw(-0.5,0.7)coordinate(A) node[left] {$U_{b}$};
			\draw(-0.5,-0.7)coordinate(A) node {$\bullet$};
			\draw(-0.5,-0.7)coordinate(A) node[left] {$U_{a}^\dagger$};
			
		}
		%\end{equation*}
	\end{equation*}
	\caption{Diagrammatic representation of $|\Omega(b)\rangle_{AB}\langle\Omega(a)|$.}
	\label{fig: outer product}
\end{figure}
The trace operation on the left side creates the inner product $\langle\Omega(a)|\Omega(b)\rangle$; equivalently, closing the diagram on the right side forms a loop with two black dots, making the pictorial representation of the orthonormality condition evident. When $a=b$, the diagram represents the projector $|\Omega(a)\rangle_{AB}\langle\Omega(a)|$, and summing over $a$ on both sides leads to the graphical completeness relation.

For example, consider the two-qubit Bell basis $|\phi(\alpha_k \beta_k)\rangle$ in \eqref{twoqubitbellstate}. Its orthonormality condition \eqref{two-qubit bell orthonormal condition} is illustrated in Fig.~\ref{fig: orthonormal condition}.
\begin{figure}[!hbt]
	\begin{equation*} 
		%	\begin{equation*}
			\tikz[baseline]{
				\draw(1,-1)--(1,1);
				\draw(1,1)--(2,1);
				\draw(2,1)--(2,-1);
				\draw(1,-1)--(2,-1);
				\draw[dashed](0.8,0)--(2.2,0);
				\draw(1,0.5)coordinate(A)node{$\bullet$};
				\draw(1,0.5)coordinate(A)node[left]{$T^\dagger(\alpha_k^\prime \beta_k^\prime)$};
				\draw(1,-0.5)coordinate(A)node{$\bullet$};
				\draw(1,-0.5)coordinate(A)node[left]{$T(\alpha_k \beta_k)$};
				\draw(3.5,0)coordinate(A)node{$=\delta_{\alpha_k^\prime \alpha_k } \delta_{\beta_k^\prime \beta_k} $}
			}
		\end{equation*}
		\caption{Diagrammatic representation of the orthonormality condition \eqref{two-qubit bell orthonormal condition}.}
		\label{fig: orthonormal condition}
	\end{figure}
	
	The projectors $|\phi(\alpha_k \beta_k)\rangle\langle \phi(\alpha_k \beta_k)  |$ are respectively shown in Fig.~\ref{fig: two qubit orthonormal projectors}. 
	\begin{figure}[!hbt]
		\begin{equation*}
			\tikz[baseline]{
				\draw(-2,1)--(-2,0.2);
				\draw(-2,0.2)--(-1,0.2);
				\draw(-1,0.2)--(-1,1);
				\draw(-2,-1)--(-2,-0.2);
				\draw(-2,-0.2)--(-1,-0.2);
				\draw(-1,-0.2)--(-1,-1);
				\draw(2,1)--(2,0.2);
				\draw(2,0.2)--(1,0.2);
				\draw(1,0.2)--(1,1);
				\draw(2,-1)--(2,-0.2);
				\draw(2,-0.2)--(1,-0.2);
				\draw(1,-0.2)--(1,-1);
				\draw(1,0.6)coordinate(A)node{$\bullet$};
				\draw(1,0.6)coordinate(A)node[right]{${Z}$};
				\draw(1,-0.6)coordinate(A)node{$\bullet$};
				\draw(1,-0.6)coordinate(A)node[right]{${Z}$};
				\draw(-1.5,-1)coordinate(A)node[below]{$|\phi(00)\rangle\langle\phi(00)|$};
				\draw(1.5,-1)coordinate(A)node[below]{$|\phi(10)\rangle\langle\phi(10)|$};
				\draw(4,1)--(4,0.2);
				\draw(4,0.2)--(5,0.2);
				\draw(5,0.2)--(5,1);
				\draw(4,-1)--(4,-0.2);
				\draw(4,-0.2)--(5,-0.2);
				\draw(5,-0.2)--(5,-1);
				\draw(7,1)--(7,0.2);
				\draw(7,0.2)--(8,0.2);
				\draw(8,0.2)--(8,1);
				\draw(7,-1)--(7,-0.2);
				\draw(7,-0.2)--(8,-0.2);
				\draw(8,-0.2)--(8,-1);
				\draw(7,0.6)coordinate(A)node{$\bullet$};
				\draw(7,0.6)coordinate(A)node[right]{${ZX}$};
				\draw(7,-0.6)coordinate(A)node{$\bullet$};
				\draw(7,-0.6)coordinate(A)node[right]{${XZ}$};
				\draw(4,0.6)coordinate(A)node{$\bullet$};
				\draw(4,0.6)coordinate(A)node[right]{${X}$};
				\draw(4,-0.6)coordinate(A)node{$\bullet$};
				\draw(4,-0.6)coordinate(A)node[right]{${X}$};
				\draw(4.5,-1)coordinate(A)node[below]{$|\phi(01)\rangle\langle\phi(01)|$};
				\draw(7.5,-1)coordinate(A)node[below]{$|\phi(11)\rangle\langle\phi(11)|$};
				
				\draw(0,0)coordinate(A)node{$;$};
				\draw(3,0)coordinate(A)node{$;$};
				\draw(6,0)coordinate(A)node{$;$};
				
			}
		\end{equation*}
		\caption{Diagrammatic representation of the projectors $|\phi(\alpha_k \beta_k)\rangle\langle \phi(\alpha_k \beta_k)|$.}
		\label{fig: two qubit orthonormal projectors}
	\end{figure}
	
\begin{figure}[!hbt]
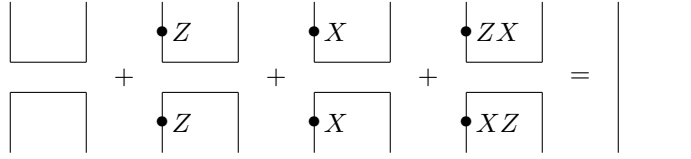

		\begin{equation*}
			%	\label{graphic two-qubit bell completeness relation}
			\tikz[baseline]{
				\draw(-2,1)--(-2,0.2);
				\draw(-2,0.2)--(-1,0.2);
				\draw(-1,0.2)--(-1,1);
				\draw(-2,-1)--(-2,-0.2);
				\draw(-2,-0.2)--(-1,-0.2);
				\draw(-1,-0.2)--(-1,-1);
				
				\draw(1,1)--(1,0.2);
				\draw(1,0.2)--(0,0.2);
				\draw(0,0.2)--(0,1);
				\draw(1,-1)--(1,-0.2);
				\draw(1,-0.2)--(0,-0.2);
				\draw(0,-0.2)--(0,-1);
				\draw(0,0.6)coordinate(A)node{$\bullet$};
				\draw(0,0.6)coordinate(A)node[right]{${Z}$};
				\draw(0,-0.6)coordinate(A)node{$\bullet$};
				\draw(0,-0.6)coordinate(A)node[right]{${Z}$};
				
				\draw(2,1)--(2,0.2);
				\draw(2,0.2)--(3,0.2);
				\draw(3,0.2)--(3,1);
				\draw(2,-1)--(2,-0.2);
				\draw(2,-0.2)--(3,-0.2);
				\draw(3,-0.2)--(3,-1);
				
				\draw(4,1)--(4,0.2);
				\draw(4,0.2)--(5,0.2);
				\draw(5,0.2)--(5,1);
				\draw(4,-1)--(4,-0.2);
				\draw(4,-0.2)--(5,-0.2);
				\draw(5,-0.2)--(5,-1);
				
				\draw(4,0.6)coordinate(A)node{$\bullet$};
				\draw(4,0.6)coordinate(A)node[right]{${ZX}$};
				\draw(4,-0.6)coordinate(A)node{$\bullet$};
				\draw(4,-0.6)coordinate(A)node[right]{${XZ}$};
				
				\draw(2,0.6)coordinate(A)node{$\bullet$};
				\draw(2,0.6)coordinate(A)node[right]{${X}$};
				\draw(2,-0.6)coordinate(A)node{$\bullet$};
				\draw(2,-0.6)coordinate(A)node[right]{${X}$};
				\draw(-0.5,0)coordinate(A)node{$+$};
				\draw(1.5,0)coordinate(A)node{$+$};
				\draw(3.5,0)coordinate(A)node{$+$};
				\draw(5.5,0)coordinate(A)node{$=$};
				\draw(6,1)--(6,-1);
				\draw(7,1)--(7,-1);
			} \,\,\,\,\,
		\end{equation*}
		\caption{Diagrammatic representation of the completeness relation \eqref{two-qubit bell completeness relation}.}
		\label{fig: graphic two-qubit bell completeness relation}
	\end{figure}

	The algebraic completeness relation \eqref{two-qubit bell completeness relation} is illustrated in Fig.~\ref{fig: graphic two-qubit bell completeness relation}.

\subsection{Diagrammatic generalized \texorpdfstring{\(2n\)}{}-qubit Bell states}

As an example, let us clarify the graphical representation of the generalized Bell state $|\mathcal{B}_{4}\rangle_{A_1A_2B_1B_2}$ in a system of four qubits $A_1, A_2, B_1, B_2$. The state
\begin{equation}
	|\mathcal{B}_{4}\rangle_{A_1A_2B_1B_2}=\tau_4 (|\phi\rangle_{A_1B_1}\otimes|\phi\rangle_{A_2B_2}),
\end{equation} 
is equal to the generalized bipartite Bell state
\begin{equation}
	|\Omega\rangle_{AB}|_{d=4}=\frac{1}{2}\sum_{i=0}^3|i\rangle_{A}\otimes|i\rangle_{B},
\end{equation}
with $A$ denoting the $A_1 A_2$ qubits and $B$ denoting the $B_1 B_2$ qubits. Although the states $|\mathcal{B}_{4}\rangle_{AB}$ and $|\Omega\rangle_{AB}|_{d=4}$ are the same, they have different pictorial representations.

\begin{figure}[!hbt]
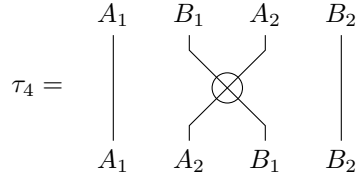

	\begin{equation*}
		%\begin{equation}	\label{twistor four} 
		\tikz[baseline]{
			\draw(1,0)coordinate(A) node {$\tau_4=$};
			\draw(2,0.7)--(2,-0.7);
			\draw(3,0.7)--(3,0.5)--(4,-0.5)--(4,-0.7);
			\draw(4,0.7)--(4,0.5)--(3,-0.5)--(3,-0.7);
			\draw(5,0.7)--(5,-0.7);
			\draw(3.5,0)circle (0.2);
			\draw(2,-0.7)coordinate(A) node[below] {$A_1$};
			\draw(2,0.7)coordinate(A) node[above] {$A_1$};
			\draw(3,-0.7)coordinate(A) node[below] {$A_2$};
			\draw(3,0.7)coordinate(A) node[above] {$B_1$};
			\draw(4,-0.7)coordinate(A) node[below] {$B_1$};
			\draw(4,0.7)coordinate(A) node[above] {$A_2$};
			\draw(5,-0.7)coordinate(A) node[below] {$B_2$};
			\draw(5,0.7)coordinate(A) node[above] {$B_2$};
		} 
		%\end{equation}
	\end{equation*}
	\caption{Diagrammatic representation of the twist operator $\tau_4$.}
	\label{fig: graphic twist operator four}
\end{figure}

The twist operator $\tau_4$ is graphically represented in Fig.~\ref{fig: graphic twist operator four}, where the central crossing with a small circle denotes the SWAP gate that exchanges qubits $A_2$ and $B_1$.  

\begin{figure}[!hbt]
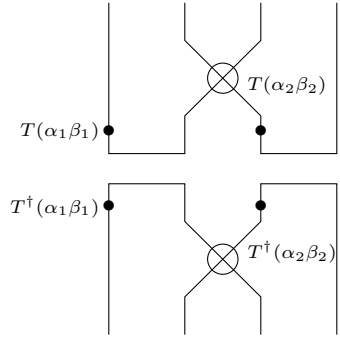

	\begin{equation*}
		%\begin{equation}	\label{projector B4}
		\tikz[baseline]{
			\draw(1,2.2)--(1,0.2)--(2,0.2)--(2,0.7)--(3,1.7)--(3,2.2);
			\draw(4,2.2)--(4,0.2)--(3,0.2)--(3,0.7)--(2,1.7)--(2,2.2);
			\draw(1,-2.2)--(1,-0.2)--(2,-0.2)--(2,-0.7)--(3,-1.7)--(3,-2.2);
			\draw(4,-2.2)--(4,-0.2)--(3,-0.2)--(3,-0.7)--(2,-1.7)--(2,-2.2);
			\draw(2.5,1.2)circle (0.2);
			\draw(2.5,-1.2)circle (0.2);
			\draw(1,0.5)coordinate(A)node{$\bullet$};
			\draw(1,0.5)coordinate(A)node[left]{$\scriptstyle T(\alpha_1\beta_1)$};
			\draw(1,-0.5)coordinate(A)node{$\bullet$};
			\draw(1,-0.5)coordinate(A)node[left]{$\scriptstyle T^\dagger(\alpha_1\beta_1)$};
			\draw(3,0.5)coordinate(A)node{$\bullet$};
			\draw(2.7,1.1)coordinate(A)node[right]{$\scriptstyle T(\alpha_2\beta_2)$};
			\draw(3,-0.5)coordinate(A)node{$\bullet$};
			\draw(2.7,-1.1)coordinate(A)node[right]{$\scriptstyle  T^\dagger(\alpha_2\beta_2)$}
		} 
		%\end{equation}
	\end{equation*}
	\caption{Diagrammatic representation of the projector $|\mathcal{B}_{4}(\underline{\alpha\beta})\rangle \langle \mathcal{B}_{4}(\underline{\alpha\beta})|$.}
	\label{fig: graphic four qubit Bell state}
\end{figure}

The diagrammatic projector $|\mathcal{B}_{4}(\underline{\alpha\beta})\rangle \langle \mathcal{B}_{4}(\underline{\alpha\beta})|$ is illustrated in Fig.~\ref{fig: graphic four qubit Bell state}, where the upper diagram without black dots simply represents the generalized Bell state $|\mathcal{B}_{4}\rangle$. 

In general, a diagrammatic generalized $2n$-qubit Bell state $|\mathcal{B}_{2n}(\underline{\alpha\beta})\rangle$ is obtained by applying the graphical twist operator $\tau_{2n}$ to the pictorial $n$-fold tensor product of two-qubit Bell states $|\phi(\alpha_k \beta_k)\rangle$.

\subsection{Diagrammatic generalized multiple qubit state}

In addition to the pictorial states \(|\mathcal{B}_{2n}(\underline{\alpha\beta})\rangle\), other graphical states are needed to describe a quantum information protocol within the extended Temperley--Lieb diagrammatic approach.

\begin{figure}[!hbt]
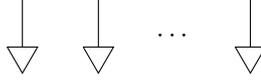

	\begin{equation*}
		% \begin{equation}	\label{nqubit1}
			\tikz[baseline]{
				\draw(1,0)--(1,-0.6536);
				\draw(0.8,-0.6536)--(1.2,-0.6536)--(1,-1)--(0.8,-0.6536);
				\draw(2,0)--(2,-0.6536);
				\draw(1.8,-0.6536)--(2.2,-0.6536)--(2,-1)--(1.8,-0.6536);
				\draw(3,-0.5)coordinate(A) node {$\cdots$};
				\draw(4,0)--(4,-0.6536);
				\draw(3.8,-0.6536)--(4.2,-0.6536)--(4,-1)--(3.8,-0.6536);
				%	\draw(4.8,-0.5)coordinate(A) node {.};
			} 
			% \end{equation}
	\end{equation*}
	\caption{Diagrammatic representation of a multiple qubit product state.}
	\label{fig: graphic multi qubit product state}
\end{figure}

A tensor product state of a multiple qubit system, such as a product basis state for the Hilbert space, is simply illustrated in Fig.~\ref{fig: graphic multi qubit product state}.

\begin{figure}[!hbt]
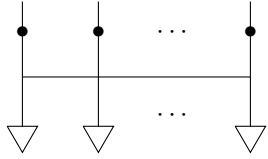

	\begin{equation*}
		%{equation}\label{nqubit2}
		\tikz[baseline]{
			\draw(3,-0.5)coordinate(A) node {$\cdots$};
			\draw(3,0.6)coordinate(A) node {$\cdots$};
			\draw(1,1)--(1,-0.6536);
			\draw(0.8,-0.6536)--(1.2,-0.6536)--(1,-1)--(0.8,-0.6536);
			\draw(2,1)--(2,-0.6536);
			\draw(1.8,-0.6536)--(2.2,-0.6536)--(2,-1)--(1.8,-0.6536);
			\draw(4,1)--(4,-0.6536);
			\draw(3.8,-0.6536)--(4.2,-0.6536)--(4,-1)--(3.8,-0.6536);
			\draw(1,0.0)--(4,0.0);
			\draw(1,0.6)coordinate(A) node {$\bullet$};
			\draw(2,0.6)coordinate(A) node {$\bullet$};
			\draw(4,0.6)coordinate(A) node {$\bullet$};
			
		} 
		%  \end{equation}
\end{equation*}
\caption{Diagrammatic representation of a multiple qubit entangled state.}
\label{fig: graphic multi qubit entangled state}
\end{figure}

A schematic representation of a generalized multiple qubit entangled state is shown in Fig.~\ref{fig: graphic multi qubit entangled state}, with black dots signifying the local action of quantum gates on qubits.

Let us now focus on diagrammatic quantum teleportation using cup and cap states.

\section{Diagrammatic teleportation via generalized Bell states}

Since the teleportation equation \eqref{teleportation equation 2} encapsulates all the essential information for a complete teleportation protocol \cite{Bennett_1993} to transmit an unknown qubit, the corresponding teleportation equations based on generalized Bell states clearly demonstrate how to transport an unknown single qudit or multiple qubits in a step-by-step manner.

In this section, we derive teleportation equations using generalized two-qudit Bell states and generalized multiple qubit Bell states, respectively. We also provide a pictorial proof of the teleportation equation \eqref{teleportation equation 2} to illustrate the method within the extended Temperley--Lieb diagrammatic approach. Finally, we present a visual description of quantum teleportation of a single qudit or multiple qubits.

\subsection{Teleportation equations via generalized two-qudit Bell states}

Given the orthonormal basis $|i\rangle$ for a $d$-dimensional Hilbert space $\mathscr{H}_d$, a qudit state $|\psi_d\rangle$ is expressed as $|\psi_d\rangle=\sum_{i=0}^{d-1} c_i|i\rangle$ with amplitudes $c_i$ satisfying the normalization condition $\sum_i|c_i|^2=1$. 

Alice has an unknown qudit $|\psi_d\rangle$ and wants to transfer it to Bob at a distant location. To achieve this, Alice and Bob share a generalized two-qudit Bell state $|\Omega M^{\intercal}(b)\rangle$, and the reformulation of the prepared state yields the teleportation equation
\begin{equation}
	\label{qudit teleportation equation11}
	|\psi_d\rangle \otimes |\Omega M^{\intercal}(b) \rangle
	=\frac{1}{d}\sum_{a=0}^{d^2-1}|\Omega(a)\rangle \otimes M U_b^\intercal U_a^\dagger|\psi_d\rangle
\end{equation} 
which is shown in Fig.~\ref{fig: graphic teleportation equation qudit 11} within the extended Temperley--Lieb diagrammatic approach.
\begin{figure}[!hbt]
	\begin{equation*}
		\tikz[baseline]
		{
			\draw(-3,2)--(-3,-0.6536);
			
			\draw(-3.2,-0.6536)--(-2.8,-0.6536);
			\draw(-3.2,-0.6536)--(-3,-1);
			\draw(-2.8,-0.6536)--(-3,-1);
			
			\draw(-2,2)--(-2,-1);
			\draw(-2,1.0)coordinate(A)node{$\bullet$};
			\draw(-2,1.0)coordinate(A)node[left]{$U_b$};
			
			\draw(-2,-1)--(-1,-1);
			
			\draw(-1,2)--(-1,-1);
			\draw(-1,1.0)coordinate(A)node{$\bullet$};
			\draw(-1,1.0)coordinate(A)node[right]{$M$};
			
			\draw(0.0,0.4)coordinate(A)node{${\displaystyle =\frac{1}{d}\sum_{a=0}^{d^2-1}}$};
			
			\draw(1,2)--(1,1.2);
			\draw(1,1.2)--(2,1.2);
			\draw(2,1.2)--(2,2);
			
			\draw(3,2)--(3,-0.6536);
			
			\draw(3.2,-0.6536)--(2.8,-0.6536);
			\draw(3.2,-0.6536)--(3,-1);
			\draw(2.8,-0.6536)--(3,-1);
			
			\draw(1,1.6)coordinate(A)node{$\bullet$};
			\draw(1,1.6)coordinate(A)node[left]{$U_a$};
			
			\draw(3,1.5)coordinate(A)node{$\bullet$};
			\draw(3,1.5)coordinate(A)node[right]{$M$};
			\draw(3,1.0)coordinate(A)node{$\bullet$};
			\draw(3,1.0)coordinate(A)node[right]{$U_b^\intercal$};
			\draw(3,.5)coordinate(A)node{$\bullet$};
			\draw(3,.5)coordinate(A)node[right]{$U^\dagger_a$};
		}
	\end{equation*}
	\caption{Diagrammatic representation of the teleportation equation \eqref{qudit teleportation equation11}.}
	\label{fig: graphic teleportation equation qudit 11}
\end{figure}

For the special case $M=U_b=1\!\!1_d$, the teleportation equation \eqref{qudit teleportation equation11} simplifies to
\begin{equation}
	|\psi_d\rangle \otimes |\Omega\rangle
	=\frac{1}{d}\sum_{a=0}^{d^2-1}|\Omega(a)\rangle \otimes U_a^\dagger|\psi_d\rangle.  
\end{equation} 
This equation describes all the necessary steps for faithful teleportation of an unknown qudit state $|\psi_d\rangle$, including state preparation, generalized Bell measurements, classical communication, and unitary corrections, as in the standard teleportation of an unknown qubit \cite{Bennett_1993}. 

With $U_b=1\!\!1_d$ and an arbitrary $d\times d$ matrix $M$, the teleportation equation \eqref{qudit teleportation equation11} becomes 
\begin{equation}
	|\psi_d\rangle \otimes|\mathcal{M}^\intercal\rangle
	=\frac{1}{d}\sum_{a=0}^{d^2-1}|\Omega(a)\rangle \otimes M U_a^\dagger|\psi_d\rangle.  
\end{equation} 
When $M$ is unitary, Alice and Bob share the maximally entangled bipartite pure state $|\mathcal{M}^\intercal\rangle$. The standard teleportation procedure remains the same, except that Bob applies the unitary correction operators $U_a M^\dagger$ to his qudit in the final step.

As noted above, a generalized Bell measurement is specified by the projectors $|\Omega(a)\rangle \langle \Omega(a)|$. It can also be implemented using the projectors $|\Omega M^{\intercal}(a) \rangle \langle \Omega M^{\intercal}(a) |$ with a unitary matrix $M$. Accordingly, another teleportation equation is given by
\begin{equation}
	\label{qudit teleportation equation22}
	|\psi_d\rangle \otimes |M \Omega (b) \rangle
	=\frac{1}{d}\sum_{a=0}^{d^2-1}|\Omega M^{\intercal}(a) \rangle \otimes  U_b^\intercal U_a^\dagger|\psi_d\rangle
\end{equation} 
which is illustrated in Fig.~\ref{fig: graphic teleportation equation qudit 22}.
\begin{figure}[!hbt]
	\begin{equation*}
		\tikz[baseline]
		{
			\draw(-3,2)--(-3,-0.6536);
			
			\draw(-3.2,-0.6536)--(-2.8,-0.6536);
			\draw(-3.2,-0.6536)--(-3,-1);
			\draw(-2.8,-0.6536)--(-3,-1);
			
			\draw(-2,2)--(-2,-1);
			\draw(-2,.5)coordinate(A)node{$\bullet$};
			\draw(-2,.5)coordinate(A)node[left]{$U_b$};
			\draw(-2,1.0)coordinate(A)node{$\bullet$};
			\draw(-2,1.0)coordinate(A)node[left]{$M$};
			
			\draw(-2,-1)--(-1,-1);
			
			\draw(-1,2)--(-1,-1);
			
			\draw(0.0,0.4)coordinate(A)node{${\displaystyle =\frac{1}{d}\sum_{a=0}^{d^2-1}}$};
			
			\draw(1,2)--(1,1.2);
			\draw(1,1.2)--(2,1.2);
			\draw(2,1.2)--(2,2);
			
			\draw(3,2)--(3,-0.6536);
			
			\draw(3.2,-0.6536)--(2.8,-0.6536);
			\draw(3.2,-0.6536)--(3,-1);
			\draw(2.8,-0.6536)--(3,-1);
			
			\draw(1,1.6)coordinate(A)node{$\bullet$};
			\draw(1,1.6)coordinate(A)node[left]{$U_a$};
			
			\draw(2,1.6)coordinate(A)node{$\bullet$};
			\draw(2,1.6)coordinate(A)node[right]{$M$};
			
			\draw(3,1.0)coordinate(A)node{$\bullet$};
			\draw(3,1.0)coordinate(A)node[right]{$U_b^\intercal$};
			\draw(3,.5)coordinate(A)node{$\bullet$};
			\draw(3,.5)coordinate(A)node[right]{$U^\dagger_a$};
		}
	\end{equation*}
	\caption{Diagrammatic representation of the teleportation equation \eqref{qudit teleportation equation22}.}
	\label{fig: graphic teleportation equation qudit 22}
\end{figure}

When the basis group is generated by generalized Pauli matrices, the generalized two-qudit Bell states are denoted by $|\Omega(\alpha\beta)\rangle$, with projective measurement operators $|\Omega(\alpha\beta)\rangle \langle \Omega(\alpha\beta)|$. In this case, the teleportation equation \eqref{qudit teleportation equation11} becomes
\begin{equation}
	\label{qudit teleportation equation11p}
	|\psi_d\rangle \otimes |\Omega M^{\intercal}(a^\prime b^\prime) \rangle
	=\frac{1}{d}\sum_{\alpha,\beta=0}^{d-1}|\Omega(\alpha\beta)\rangle \otimes M U_{a^\prime b^\prime}^\intercal U_{\alpha\beta}^\dagger|\psi_d\rangle,
\end{equation} 
where $U_{a^\prime b^\prime}^\intercal \neq U_{a^\prime b^\prime}^\dagger$ since $\mathcal{Z}^\intercal \neq \mathcal{Z}^\dagger$. Similarly, the teleportation equation \eqref{qudit teleportation equation22} is reformulated as
\begin{equation}
	\label{qudit teleportation equation22p}
	|\psi_d\rangle \otimes |M \Omega (a^\prime b^\prime) \rangle
	=\frac{1}{d}\sum_{\alpha,\beta=0}^{d-1}|\Omega M^{\intercal}(\alpha\beta) \rangle \otimes  U_{a^\prime b^\prime}^\intercal U_{\alpha\beta}^\dagger|\psi_d\rangle
\end{equation} 
with a $d\times d$ unitary matrix $M$.

\subsection{Teleportation equations via generalized multiple qubit Bell states}

To study quantum teleportation of multiple qubits, an $n$-qubit state $|\underline{\psi}_n\rangle$ is first expressed as 
\begin{equation}
	|\underline{\psi}_n\rangle=\sum_{i_1,\dots,i_n=0}^1 c_{i_1i_2\cdots i_n}|i_1 i_2\cdots i_n\rangle,
\end{equation}
or simply denoted as $|\underline{\psi}_n\rangle=\sum_{\underline{i}} c_{\underline{i}}|\underline{i}\rangle$,
with the normalization condition $\sum_{\underline{i}} |c_{\underline{i}}|^2=1$. Given the generalized projective Bell measurement operators $|\mathcal{B}_{2n}(\underline{\alpha\beta})\rangle \langle \mathcal{B}_{2n}(\underline{\alpha\beta})|$, Alice and Bob share the generalized $2n$-qubit Bell state $|\mathcal{B} M^\intercal_{2n}(\underline{a^\prime b^\prime})\rangle$, and the teleportation equation takes the form
\begin{equation}
	\label{n-qubit teleportation equation11}
	|\underline{\psi}_n\rangle \otimes |\mathcal{B} M^\intercal_{2n}(\underline{a^\prime b^\prime}) \rangle
	=\frac {1} {2^n} \sum_{\underline{\alpha},\underline{\beta}}|\mathcal{B}_{2n}(\underline{\alpha\beta})\rangle \otimes M T_n^\dagger(\underline{a^\prime b^\prime}) T_n^\dagger(\underline{\alpha\beta}) |\underline{\psi}_n\rangle,
\end{equation} 
with an $n$-qubit quantum gate $M$ and $n$-bit strings $\underline{a}^\prime$, $\underline{b}^\prime$, $\underline{\alpha}$, and $\underline{\beta}$. However, if Alice uses the generalized projective Bell measurement operators $|\mathcal{B} M^\intercal_{2n}(\underline{\alpha\beta})\rangle \langle \mathcal{B} M^\intercal_{2n}(\underline{\alpha\beta})|$, then the teleportation equation becomes
\begin{equation}
	\label{n-qubit teleportation equation22}
	|\underline{\psi}_n\rangle \otimes |M\mathcal{B}_{2n}(\underline{a^\prime b^\prime}) \rangle
	=\frac {1} {2^n} \sum_{\underline{\alpha},\underline{\beta}}|\mathcal{B} M^\intercal_{2n}(\underline{\alpha\beta})\rangle \otimes T_n^\dagger(\underline{a^\prime b^\prime}) T_n^\dagger(\underline{\alpha\beta}) |\underline{\psi}_n\rangle,
\end{equation} 
with $T_n^\dagger(\underline{a^\prime b^\prime})=T_n^\intercal(\underline{a^\prime b^\prime})$. 

Examining the teleportation equations above, we observe that they depend linearly on the unknown qudit state $|\psi_d\rangle$ (or multiple qubit state $|\underline{\psi}_n\rangle$). Therefore, it is sufficient to present the teleportation equations using only the basis vectors $|i\rangle$ (or $|\underline{i}\rangle$) instead of the full states $|\psi_d\rangle$ (or $|\underline{\psi}_n\rangle$) in algebraic calculations. Consequently, in the diagrammatic description of teleportation of an unknown $n$-qubit state, it is enough to draw the diagrammatic product state $|\underline{i}\rangle$ from Fig.~\ref{fig: graphic multi qubit product state} rather than the entangled state $|\underline{\psi}_n\rangle$ from Fig.~\ref{fig: graphic multi qubit entangled state}.

\subsection{Diagrammatic proof of the teleportation equation 
	\texorpdfstring{\eqref{teleportation equation 2}}{}}

As a matter of fact, all the above teleportation equations can be derived algebraically with little effort. However, it is more instructive to establish them through a series of visual operations within the extended Temperley--Lieb diagrammatic approach. Let us focus on a graphical proof of the teleportation equation \eqref{teleportation equation 2} as a typical example.

\begin{figure}[!hbt]
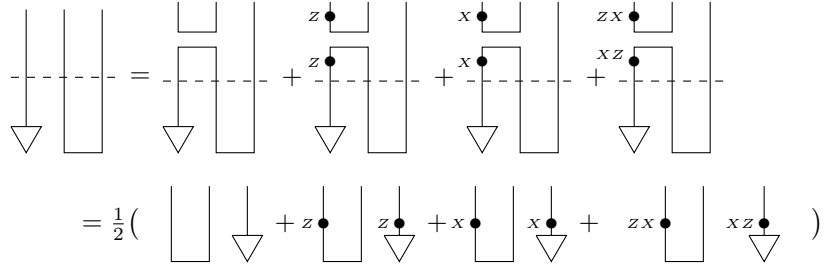

	\begin{equation*}
		\begin{aligned}
			&\tikz[baseline]{
				\draw[dashed](0.8,0)--(2.2,0);
				\draw(1,0.9)--(1,-0.6536);
				\draw(0.8,-0.6536)--(1.2,-0.6536)--(1,-1)--(0.8,-0.6536);
				\draw(1.5,0.9)--(1.5,-1)--(2,-1)--(2,0.9);
				
				\draw(2.5,0)coordinate (A) node {$=$};
				
				\draw(3,1)--(3,0.6)--(3.5,0.6)--(3.5,1);
				\draw(2.8,-0.6536)--(3.2,-0.6536)--(3,-1)--(2.8,-0.6536);
				\draw(3,-0.6536)--(3,0.4)--(3.5,0.4)--(3.5,-1)--(4,-1)--(4,1); 
				\draw[dashed](2.8,-0.05)--(4.2,-0.05);
				
				\draw(4.5,0)coordinate (A) node {$+$};
				
				\draw(5,1)--(5,0.6)--(5.5,0.6)--(5.5,1);
				\draw(4.8,-0.6536)--(5.2,-0.6536)--(5,-1)--(4.8,-0.6536);
				
				\draw(5,-0.6536)--(5,0.4)--(5.5,0.4)--(5.5,-1)--(6,-1)--(6,1); 
				\draw[dashed](4.8,-0.05)--(6.2,-0.05);
				\draw(5,0.8)coordinate(A) node {$\bullet$};
				\draw(5,0.8)coordinate(A) node[left] {$\scriptscriptstyle Z$};
				\draw(5,0.2)coordinate(A) node {$\bullet$};
				\draw(5,0.2)coordinate(A) node[left] {$\scriptscriptstyle Z$};
				
				\draw(6.5,0)coordinate (A) node {$+$};
				
				\draw(7,1)--(7,0.6)--(7.5,0.6)--(7.5,1);
				\draw(7,-0.6536)--(7,0.4)--(7.5,0.4)--(7.5,-1)--(8,-1)--(8,1); 
				\draw(6.8,-0.6536)--(7.2,-0.6536)--(7,-1)--(6.8,-0.6536);
				\draw[dashed](6.8,-0.05)--(8.2,-0.05);
				\draw(7,0.8)coordinate(A) node {$\bullet$};
				\draw(7,0.8)coordinate(A) node[left] {$\scriptscriptstyle X$};
				\draw(7,0.2)coordinate(A) node {$\bullet$};
				\draw(7,0.2)coordinate(A) node[left] {$\scriptscriptstyle X$};
				
				\draw(8.5,0)coordinate (A) node {$+$};
				
				\draw(9,1)--(9,0.6)--(9.5,0.6)--(9.5,1);
				\draw(9,-0.6536)--(9,0.4)--(9.5,0.4)--(9.5,-1)--(10,-1)--(10,1); 
				\draw(8.8,-0.6536)--(9.2,-0.6536)--(9,-1)--(8.8,-0.6536);
				\draw[dashed](8.8,-0.05)--(10.2,-0.05);
				\draw(9,0.8)coordinate(A) node {$\bullet$};
				\draw(9,0.8)coordinate(A) node[left] {$\scriptscriptstyle ZX$};
				\draw(9,0.2)coordinate(A) node {$\bullet$};
				\draw(9,0.3)coordinate(A) node[left] {$\scriptscriptstyle XZ$};
				
			}\\
			&\tikz[baseline]{
				\draw(1,0)coordinate(A) node {};
				\draw(2.2,-0.5)coordinate (A) node {$=\frac{1}{2} \big( $};
				\draw(3,0)--(3,-1)--(3.5,-1)--(3.5,0);
				\draw(4,0)--(4,-0.6536);
				\draw(3.8,-0.6536)--(4.2,-0.6536)--(4,-1)--(3.8,-0.6536);
				\draw(4.5,-0.5)coordinate(A) node {$+$};
				\draw(5,0)--(5,-1)--(5.5,-1)--(5.5,0);
				\draw(6,0)--(6,-0.6536);
				\draw(5.8,-0.6536)--(6.2,-0.6536)--(6,-1)--(5.8,-0.6536);
				\draw(5,-0.5)coordinate(A) node {$\bullet$};
				\draw(5,-0.5)coordinate(A) node[left] {$\scriptscriptstyle Z$};
				\draw(6,-0.5)coordinate(A) node {$\bullet$};
				\draw(6,-0.5)coordinate(A) node[left] {$\scriptscriptstyle Z$};
				\draw(6.5,-0.5)coordinate(A) node {$+$};
				\draw(7,0)--(7,-1)--(7.5,-1)--(7.5,0);
				\draw(8,0)--(8,-0.6536);
				\draw(7.8,-0.6536)--(8.2,-0.6536)--(8,-1)--(7.8,-0.6536);
				\draw(7,-0.5)coordinate(A) node {$\bullet$};
				\draw(7,-0.5)coordinate(A) node[left] {$\scriptscriptstyle X$};
				\draw(8,-0.5)coordinate(A) node {$\bullet$};
				\draw(8,-0.5)coordinate(A) node[left] {$\scriptscriptstyle X$};
				\draw(8.5,-0.5)coordinate(A) node {$+$};
				\draw(9.5,0)--(9.5,-1)--(10,-1)--(10,0);
				\draw(10.8,0)--(10.8,-0.6536);
				\draw(10.6,-0.6536)--(11,-0.6536)--(10.8,-1)--(10.6,-0.6536);
				\draw(9.5,-0.5)coordinate(A) node {$\bullet$};
				\draw(9.5,-0.5)coordinate(A) node[left] {$\scriptscriptstyle ZX$};
				\draw(10.8,-0.5)coordinate(A) node {$\bullet$};
				\draw(10.8,-0.5)coordinate(A) node[left] {$\scriptscriptstyle XZ$};
				\draw(11.5,-0.5)coordinate(A) node {$\big)$};
			}
		\end{aligned}
	\end{equation*}
	\caption{Diagrammatic proof of the teleportation equation \eqref{teleportation equation 2}.}
	\label{fig: graphic proof teleportation equation qubit}
\end{figure}

Examining the teleportation equation \eqref{qudit teleportation equation11} and its diagrammatic representation, it is evident that the chosen teleportation equation \eqref{teleportation equation 2} and its pictorial description are obtained by setting \(M=U_b=1\!\!1_2\) and requiring that the basis group be generated by the Pauli matrices. The corresponding schematic procedure is clearly explained in Fig.~\ref{fig: graphic proof teleportation equation qubit}.

To derive the diagrammatic teleportation equation shown in Fig.~\ref{fig: graphic proof teleportation equation qubit}, at least three crucial steps are required.
First, the graphical completeness relation for the projective measurement operators onto two-qubit Bell states, shown in Fig.~\ref{fig: graphic two-qubit bell completeness relation}, is applied to the left-hand side of the teleportation equation \eqref{teleportation equation 2}. Specifically, the part belonging to the first and second Hilbert spaces above the dashed line in Fig.~\ref{fig: graphic proof teleportation equation qubit} corresponds to the pictorial completeness relation in Fig.~\ref{fig: graphic two-qubit bell completeness relation}. Second, using the rules for moving local quantum gates on cup and cap states from one Hilbert space to the other, as shown in Fig.~\ref{fig: Rule 3}, the local Pauli gates on the lower cap are shifted from the first Hilbert space to the third. Third, via the topological-like operations associated with the transfer operator in Figs.~\ref{fig: Rule 2} and \ref{fig: Rule 4}, an upper cap together with a lower cup is deformed into an oblique line, thereby diagrammatically transporting the unknown qubit from the first Hilbert space to the third.

  \subsection{Diagrammatic teleportation of a single qudit}
  
\begin{figure}[!hbt]
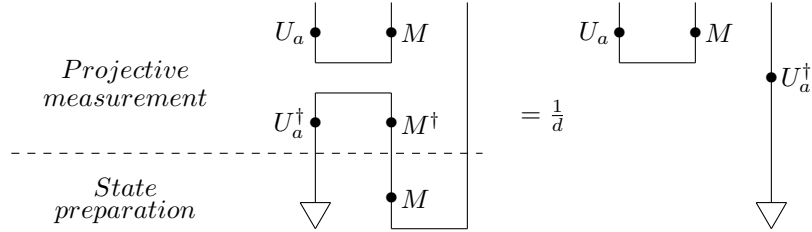

  	\begin{equation*}
  		%\label{2.23}
  		\tikz[baseline]{
  			\draw(-3,2)--(-3,1.2);
  			\draw(-3,1.2)--(-2,1.2);
  			\draw(-2,1.2)--(-2,2);
  			\draw(-3,0.8)--(-2,0.8);
  			\draw(-3,0.8)--(-3,-0.6536);
  			\draw(-3.2,-0.6536)--(-2.8,-0.6536);
  			\draw(-3.2,-0.6536)--(-3,-1);
  			\draw(-2.8,-0.6536)--(-3,-1);
  			\draw(-2,0.8)--(-2,-1);
  			\draw(-2,-1)--(-1,-1);
  			\draw(-1,-1)--(-1,2);
  			\draw(0,0.5)coordinate(A)node{$=\frac{1}{d}$};
  			\draw(1,2)--(1,1.2);
  			\draw(1,1.2)--(2,1.2);
  			\draw(2,1.2)--(2,2);
  			\draw(3,2)--(3,-0.6536);
  			\draw(3.2,-0.6536)--(2.8,-0.6536);
  			\draw(3.2,-0.6536)--(3,-1);
  			\draw(2.8,-0.6536)--(3,-1);
  			\draw(-3,1.6)coordinate(A)node{$\bullet$};
  			\draw(-3,1.6)coordinate(A)node[left]{$U_a$};
  			\draw(-2,1.6)coordinate(A)node{$\bullet$};
  			\draw(-2,1.6)coordinate(A)node[right]{$M$};
  			\draw(-3,0.4)coordinate(A)node{$\bullet$};
  			\draw(-3,0.4)coordinate(A)node[left]{$U_{a}^\dagger$};
  			\draw(-2,0.4)coordinate(A)node{$\bullet$};
  			\draw(-2,0.4)coordinate(A)node[right]{$M^\dagger$};
  			\draw(-2,-0.6)coordinate(A)node{$\bullet$};
  			\draw(-2,-0.6)coordinate(A)node[right]{$M$};
  			\draw(1,1.6)coordinate(A)node{$\bullet$};
  			\draw(1,1.6)coordinate(A)node[left]{$U_{a}$};
  			\draw(2,1.6)coordinate(A)node{$\bullet$};
  			\draw(2,1.6)coordinate(A)node[right]{$M$};
  			\draw(3,1)coordinate(A)node{$\bullet$};
  			\draw(3,1)coordinate(A)node[right]{$U_{a}^\dagger$};
  			\draw[dashed](-7,0)--(-0.8,0);
  			\draw(-5.5,0.8)coordinate(A)node[above]{Projective};
  			\draw(-5.5,0.5)coordinate(A)node[above]{measurement};
  			\draw(-5.5,-0.2)coordinate(A)node[below]{State};
  			\draw(-5.5,-0.5)coordinate(A)node[below]{preparation};
  		}
  	\end{equation*}
  	\caption{Diagrammatic representation of the projective teleportation equation \eqref{projective qudit teleportation equation22}.}
  	\label{fig: graphic projective teleportation equation22}
  \end{figure}

  As a representative example of diagrammatic teleportation of a single qudit, we further examine the teleportation equation \eqref{qudit teleportation equation22} with \(U_b=1\!\!1_d\). First, instead of a set of observables, we use a set of orthogonal projectors \(|\Omega M^{\intercal}(a) \rangle \langle \Omega M^{\intercal}(a) |\) to characterize the quantum measurement. The projective teleportation equation is then introduced to describe the protocol using these projective measurement operators:
  \begin{equation}
  	\label{projective qudit teleportation equation22}
  	(|\Omega M^{\intercal}(a) \rangle \langle \Omega M^{\intercal}(a) |\otimes 1\!\!1_d)	
  	(|\psi_d\rangle \otimes |M \Omega \rangle)
  	=\frac{1}{d}(|\Omega M^{\intercal}(a) \rangle \otimes  U_a^\dagger|\psi_d\rangle),
  \end{equation} 
  which is illustrated in Fig.~\ref{fig: graphic projective teleportation equation22}.  
  
  Fig.~\ref{fig: graphic projective teleportation equation22} clearly illustrates the complete procedure for transmitting an unknown qudit state \(|\psi_d\rangle\) from Alice to Bob. Alice possesses the qudit state \(|\psi_d\rangle\) and shares a maximally entangled bipartite state \(|M \Omega \rangle\) with Bob; the prepared state is thus \(|\psi_d\rangle \otimes |M \Omega \rangle\). Alice then performs the projective measurement operators \(|\Omega M^{\intercal}(a) \rangle \langle \Omega M^{\intercal}(a) |\) on her two qudits and communicates her measurement results to Bob via a classical channel. Finally, Bob applies the local unitary correction operator \(U_a\) to his qudit to recover the unknown state \(|\psi_d\rangle\). Note that the case \(M=1\!\!1_d\) corresponds to the standard quantum teleportation protocol using the projective measurement operators \(|\Omega(a) \rangle \langle \Omega(a) |\).

\subsection{Diagrammatic teleportation of multiple qubits}

In view of the projective teleportation equation \eqref{projective qudit teleportation equation22}, the relevant orthogonal projectors are applied to the teleportation equations \eqref{n-qubit teleportation equation11} and \eqref{n-qubit teleportation equation22}, respectively, in order to derive the projective teleportation equations for quantum teleportation of $n$ qubits. For simplicity, we consider only the following projective teleportation equation:
\begin{equation}
	\label{projective n-qubit teleportation equation11}
	(|\mathcal{B}_{2n}(\underline{\alpha\beta})\rangle \langle \mathcal{B}_{2n}(\underline{\alpha\beta})|
	\otimes 1\!\! 1_{2^n})
	(|\underline{\psi}_n\rangle \otimes |\mathcal{B}_{2n} \rangle)
	=\frac {1} {2^n} (|\mathcal{B}_{2n}(\underline{\alpha\beta})\rangle \otimes 	T_n^\dagger(\underline{\alpha\beta}) |\underline{\psi}_n\rangle).
\end{equation} 
A schematic diagram for the teleportation of two unknown qubits ($n=2$) is shown in Fig.~\ref{diagrammatic projective 2-qubit teleportation}, where the diagrammatic twist operator $\tau_4$ from Fig.~\ref{fig: graphic twist operator four} and the diagrammatic generalized four-qubit Bell state projector from Fig.~\ref{fig: graphic four qubit Bell state} are used, along with the diagrammatic two-qubit product state from Fig.~\ref{fig: graphic multi qubit product state} instead of the diagrammatic two-qubit entangled state from Fig.~\ref{fig: graphic multi qubit entangled state}.
\begin{figure}[!hbt]
	\begin{equation*}
		%\label{105}
		\scalebox{0.8}[0.9]
		{
			\tikz[baseline]{
				\draw(1,4.5)--(1,2.5)--(2,2.5)--(2,3)--(3,4)--(3,4.5);
				\draw(4,4.5)--(4,2.5)--(3,2.5)--(3,3)--(2,4)--(2,4.5);
				\draw(1,-1.6536)--(1,2.1)--(2,2.1)--(2,1.6)--(3,0.6)--(3,-0.1);
				\draw(4,0.1)--(4,2.1)--(3,2.1)--(3,1.6)--(2,0.6)--(2,-1.6536);
				
				\draw(0.8,-1.6536)--(1.2,-1.6536)--(1,-2)--(0.8,-1.6536);
				
				\draw(1.8,-1.6536)--(2.2,-1.6536)--(2,-2)--(1.8,-1.6536);
				
				\draw(3,-0.1)--(3,-2)--(4,-2)--(4,-1.6)--(5,-0.6)--(5,4.5);
				\draw(6,4.5)--(6,-2)--(5,-2)--(5,-1.6)--(4,-0.6)--(4,0.1);
				\draw(2.5,3.5)circle (0.2);
				\draw(2.5,1.1)circle (0.2);
				\draw(4.5,-1.1)circle (0.2);
				
				\draw(7,1.25)coordinate(A)node{\scalebox{1.5}[1.5]{$=\frac{1}{4}$}};
				\draw(8,4.5)--(8,2.5)--(9,2.5)--(9,3)--(10,4)--(10,4.5);
				\draw(11,4.5)--(11,2.5)--(10,2.5)--(10,3)--(9,4)--(9,4.5);
				\draw(9.5,3.5)circle (0.2);
				\draw(12,4.5)--(12,-1.6536);
				\draw(11.8,-1.6536)--(12.2,-1.6536)--(12,-2)--(11.8,-1.6536);
				\draw(13,4.5)--(13,-1.6536);
				\draw(12.8,-1.6536)--(13.2,-1.6536)--(13,-2)--(12.8,-1.6536);
				\draw(1,2.8)coordinate(A)node{$\bullet$};
				\draw(1,2.8)coordinate(A)node[left]{$\scriptstyle T(\alpha_1\beta_1)$};
				\draw(1,1.6)coordinate(A)node{$\bullet$};
				\draw(1,1.6)coordinate(A)node[left]{$\scriptstyle T^\dagger(\alpha_1\beta_1)$};
				
				\draw(3,2.8)coordinate(A)node{$\bullet$};
				\draw(2.7,3.4)coordinate(A)node[right]{$\scriptstyle T(\alpha_2\beta_2)$};
				\draw(3,1.6)coordinate(A)node{$\bullet$};
				\draw(2.7,1.0)coordinate(A)node[right]{$\scriptstyle T^\dagger(\alpha_2\beta_2)$};
				\draw(8,2.8)coordinate(A)node{$\bullet$};
				\draw(8,2.8)coordinate(A)node[left]{$\scriptstyle T(\alpha_1\beta_1)$};
				\draw(10,2.8)coordinate(A)node{$\bullet$};
				\draw(9.7,3.4)coordinate(A)node[right]{$\scriptstyle T(\alpha_2\beta_2)$};
				\draw(12,0.6)coordinate(A)node{$\bullet$};
				\draw(12,0.6)coordinate(A)node[left]{$\scriptstyle T^\dagger(\alpha_1\beta_1)$};
				\draw(13,0.6)coordinate(A)node{$\bullet$};
				\draw(13,0.6)coordinate(A)node[right]{$\scriptstyle T^\dagger(\alpha_2\beta_2)$};
				\draw(1,-2)coordinate(A)node[below]{$A_1$};
				\draw(2,-2)coordinate(A)node[below]{$A_2$};
				\draw(3,-2)coordinate(A)node[below]{$A_1$};
				\draw(4,-2)coordinate(A)node[below]{$A_2$};
				\draw(5,-2)coordinate(A)node[below]{$B_1$};
				\draw(6,-2)coordinate(A)node[below]{$B_2$};
				\draw(8,-2)coordinate(A)node[below]{$A_1$};
				\draw(9,-2)coordinate(A)node[below]{$A_2$};
				\draw(10,-2)coordinate(A)node[below]{$A_1$};
				\draw(11,-2)coordinate(A)node[below]{$A_2$};
				\draw(12,-2)coordinate(A)node[below]{$B_1$};
				\draw(13,-2)coordinate(A)node[below]{$B_2$};
			}
		}
	\end{equation*}
	\caption{Diagrammatic representation of the teleportation equation \eqref{projective n-qubit teleportation equation11} for \(n=2\).}
	\label{diagrammatic projective 2-qubit teleportation}
\end{figure}

We now make some remarks on Fig.~\ref{diagrammatic projective 2-qubit teleportation}. First, the unknown two-qubit state $|\underline{\psi}_2\rangle$ may be taken as a product state because the teleportation equation is linear in $|\underline{\psi}_2\rangle$. Second, the most distinctive feature is the illustration of the twist operator $\tau_4$, in which the graphical SWAP gate plays a key role. Third, quantum teleportation of multiple qubits is algebraically equivalent to quantum teleportation of a single qudit, although they have different visual representations.

In general, a diagrammatic generalized $2n$-qubit Bell state $|\mathcal{B}_{2n}(\underline{\alpha\beta})\rangle$ is composed of diagrammatic SWAP gates and two-qubit Bell states. Since the generators of the Brauer algebra \cite{Brauer1937} consist of Temperley--Lieb idempotents and permutations (SWAPs), the pictorial characterization of generalized $2n$-qubit Bell states essentially provides a beautiful illustration of the Brauer algebra. It is therefore most natural to describe diagrammatic teleportation of multiple qubits within the extended Brauer diagrammatic approach, which combines the diagrammatic representation of the Brauer algebra with graphical quantum gates and quantum states.

\section{Braid teleportation via generalized Bell states}

Braid teleportation \cite{Zhang2006,Zhang2009} is defined as a reformulation of the teleportation equation in terms of the braid group representation \cite{Kauffman2012}. In this section, we first introduce the Bell transform \cite{ZZ2014} (also known as the Yang--Baxter gate \cite{KL2004}) and construct a representation of the braid group. We then review the braid teleportation of a single qubit, and finally investigate the braid teleportation of multiple qubits via generalized Bell states.

\subsection{The Bell transform, Yang--Baxter gate, and braid representation}

The Bell transform is defined as a unitary basis transformation from the product basis $|ij\rangle$ to the Bell basis $|\phi(i,j)\rangle \equiv |\phi(ij)\rangle$; it is therefore a two-qubit gate. We focus on the Bell transform \cite{ZZ2014},
\begin{equation}\label{37}
	B(\epsilon,\eta) =\frac{1}{\sqrt{2}}\begin{pmatrix}1&0&0&\eta\\0&1&\epsilon&0\\0&-\epsilon&1&0\\-\eta &0&0&1\end{pmatrix},
\end{equation}
which satisfies $B^\dagger(\epsilon,\eta)=B(-\epsilon,-\eta)$ with $\epsilon,\eta=\pm 1$. Its action on the product states yields the Bell states:
\begin{equation}
	\begin{aligned}
		B(-1,-1)|ij\rangle &=(-1)^i|\phi(i,i\oplus j)\rangle, 
		&	B(-1,1)|ij\rangle &=(-1)^{i\cdot (j\oplus 1)}|\phi(j\oplus 1,i \oplus j)\rangle,\\
		B(1,-1)|ij\rangle &=(-1)^{i\cdot j}|\phi(j,i\oplus j) \rangle,   
		&	B(1,1)|ij\rangle & =|\phi(i\oplus 1,i\oplus j)\rangle,
	\end{aligned}
\end{equation}
where binary addition and multiplication between two bits $i$ and $j$ are used.

For simplicity, the action of the Bell transform $B(\epsilon^r,\eta^r)$ with $\epsilon^r, \eta^r=\pm 1$ on the product states can be expressed in the unified form
\begin{equation}
	|\phi(i^\prime,j^\prime)\rangle=(-1)^{f(\epsilon^r,\eta^r,i,j)}B(\epsilon^r,\eta^r)|ij\rangle,
\end{equation}
where the exponents $f(\epsilon,\eta,i,j)$ are respectively given by
\begin{equation}
	\begin{aligned}
		f(-1,-1,i,j) &=i, & f(-1,1,i,j) &=i\cdot (j\oplus 1),\\ 
		f(1,-1,i,j) &=i\cdot j, & f(1,1,i,j) &=0,
	\end{aligned}
\end{equation}
and the two bits $i^\prime, j^\prime$ are a bijective function of $i, j$:
\begin{equation}
	i^\prime=i \oplus \frac{|\epsilon^r-\eta^r|}{2} j^\prime \oplus  \frac{1+\eta^r}{2}, \quad  
	j^\prime=i \oplus j,
\end{equation}
with decimal addition, subtraction, and division involved.

The Bell transform $B(\epsilon,\eta)$ is a solution of the Yang--Baxter equation \cite{Kauffman2012}
\begin{equation}\label{YBE}
	({R}\otimes1\!\!1_d)(1\!\!1_d\otimes{R})({R}\otimes1\!\!1_d)
	=(1\!\!1_d\otimes{R})({R}\otimes1\!\!1_d)(1\!\!1_d\otimes{R}),
\end{equation}
with the $R$-matrix $R=B(\epsilon,\eta)$ at $d=2$; hence $B(\epsilon,\eta)$ is the Yang--Baxter gate. 
With this, a representation of the braid group $B_n$ can be constructed as follows:
\begin{equation} 
	b_i=1\!\!1_2^{\otimes (i-1)}\otimes B(\epsilon,\eta) \otimes 1\!\!1_2^{\otimes (n-i-1)}
\end{equation}	
for $i=1, 2, \dots, n-1$, which satisfy the braid group relations  
\begin{equation} 
	b_i b_{i+1} b_i =b_{i+1} b_i b_{i+1},\quad i=1,2,\dots,n-2,
\end{equation}
and $b_i b_j =b_j b_i$ for $|i-j| > 2$. Therefore, the Yang--Baxter gate 
$B(\epsilon,\eta)$ is also called the braiding gate $B(\epsilon,\eta)$. 

In the conventional illustration of a braid group representation \cite{Kauffman2012}, the braiding gate $B(-1, 1)$ and its inverse $B(1, -1)$ have the diagrammatic representations shown in Fig.~\ref{diagrammatic braiding gates}, which will be used in quantum circuits for braid teleportation. 
\begin{figure}[!hbt]
	\begin{equation*}
		%\label{B diag rep}
		\tikz[baseline]{
			%	\draw(0.9,0.6)rectangle(2.1,-0.6);
			\draw(1,0.7)--(1,0.5)--(2,-0.5)--(2,-0.7);
			\draw(2,0.7)--(2,0.5)--(1.55,0.05);
			\draw(1.45,-0.05)--(1,-0.5)--(1,-0.7);
			\draw(1.5,-0.8)coordinate(A) node[below]{$\scriptstyle  B(-1,1)$};\quad
			\draw(2.5,0)coordinate(A) node {$;$};
			%	\draw(2.9,0.6)rectangle(4.1,-0.6);
			\draw(3,0.7)--(3,0.5)--(3.45,0.05);
			\draw(3.55,-0.05)--(4,-0.5)--(4,-0.7);
			\draw(4,0.7)--(4,0.5)--(3,-0.5)--(3,-0.7);
			\draw(3.5,-0.8)coordinate(A) node[below]{$\scriptstyle B(1,-1)$};
		}
	\end{equation*}
	\caption{Diagrammatic representations of the braiding gates $B(-1, 1)$ and $B(1, -1)$.}
	\label{diagrammatic braiding gates}
\end{figure}

Note that interested readers are referred to \cite{Zhang2006,Zhang2009, ZZP2015, ZZ2017} for further details on the Bell transform, the Yang--Baxter gate, and the braid group representation.

\subsection{Braid teleportation of a single qubit}

Let us write the teleportation equation \eqref{teleportation equation 2} in the generalized form
\begin{equation}\label{teleportation equation 1}
	|\underline{\psi}_1\rangle\otimes|\phi(k^\prime m^\prime)\rangle
	=\frac{1}{2}\sum_{i^\prime,j^\prime=0}^1|\phi(i^\prime j^\prime)
	\rangle\otimes T^\dagger(k^\prime m^\prime)  T^\dagger(i^\prime j^\prime) |\underline{\psi}_1\rangle,
\end{equation}
with two bits \(k^\prime, m^\prime=0, 1 \). In addition to the bijective mapping between the bits $i^\prime, j^\prime$ and $i, j$, there is another bijective correspondence,
\begin{equation}
	k^\prime=k \oplus \frac{|\epsilon^l-\eta^l|}{2} m^\prime \oplus  \frac{1+\eta^l}{2}, \quad  
	m^\prime=k \oplus m, 
\end{equation}
between the bits \(k^\prime, m^\prime\) and \(k, m\) with \(\epsilon^l, \eta^l=\pm 1\). The braid teleportation equation for transmitting a single qubit is obtained by reformulating the teleportation equation \eqref{teleportation equation 1} in terms of Yang--Baxter gates.

The associated braid teleportation equation then takes the form 
\begin{equation}\label{braid teleportation single qubit}
	(B(-\epsilon^r,-\eta^r)\otimes 1\!\!1_2)(1\!\!1_2\otimes B(\epsilon^l,\eta^l))
	(  |\underline{\psi}_1\rangle \otimes|km\rangle)=\frac{1}{2}\sum_{i,j=0}^1|ij\rangle\otimes U^{\epsilon^l\eta^l,\epsilon^r\eta^r}_{k^\prime m^\prime,i^\prime j^\prime}
	|\underline{\psi}_1\rangle,
\end{equation}
where the local unitary correction operators
\(U^{\epsilon^l\eta^l,\epsilon^r\eta^r}_{k^\prime m^\prime,i^\prime j^\prime}\) are given by
\begin{equation}
	U^{\epsilon^l\eta^l,\epsilon^r\eta^r}_{k^\prime m^\prime,i^\prime j^\prime}=
	(-1)^{f(\epsilon^l,\eta^l,k,m)} (-1)^{f(\epsilon^r,\eta^r,i,j)}
	T^\dagger(k^\prime m^\prime) T^\dagger(i^\prime j^\prime).
\end{equation}
Alternatively, the single-qubit gates 
\(U^{\epsilon^l\eta^l,\epsilon^r\eta^r}_{k^\prime m^\prime,i^\prime j^\prime}\) can be expressed as
\begin{equation}
	\label{single qubit gate abc}
	U^{\epsilon^l\eta^l,\epsilon^r\eta^r}_{k^\prime m^\prime,i^\prime j^\prime}=(-1)^{a}X^{b}Z^{c},
\end{equation}
with the exponents \(a, b, c\) respectively given by 
\begin{equation}
	a=f(\epsilon^l,\eta^l,k,m)\oplus f(\epsilon^r,\eta^r,i,j) \oplus k^\prime\cdot  j^\prime,
\end{equation}
and \(b=j^\prime \oplus m^\prime\) and \(c=i^\prime \oplus  k^\prime\). When \(\epsilon^r=-\epsilon^l\) and \(\eta^l=-\eta^r\), the exponents \(a, b, c\) are explicitly determined in Table~\ref{Table abc}.
\begin{table}[!hbt]
	\centering
	\begin{tabular}{c|c|c|c}
		\hline
		\hline
		$(\epsilon^l,\eta^l)$ & $a $ & $b $ & $c$        
		\\ 
		\hline
		(-1,1)            & $i \cdot j \oplus (m\oplus 1)(i\oplus j \oplus k)$     & \multirow{4}{*}{$i\oplus j \oplus k \oplus m$} & \multirow{2}{*}{$j\oplus m \oplus 1$} 
		\\ \cline{1-2}
		(1,-1)            & $ i\cdot (j\oplus 1) \oplus  m \cdot (i\oplus j\oplus k) $   &                     &                       \\ \cline{1-2} \cline{4-4}
		(1,1)             & $i \oplus (k\oplus 1)(i \oplus j)$        &                           & \multirow{2}{*}{$i \oplus k \oplus 1$}                       
		\\ \cline{1-2}
		(-1,-1)           & $ k\cdot (i\oplus j \oplus 1)$   &                     &                       \\ \hline
		\hline
	\end{tabular}  
	\caption{The single-qubit gates \eqref{single qubit gate abc} for the case \(\epsilon^r=-\epsilon^l\) and \(\eta^l=-\eta^r\).}
	\label{Table abc}
\end{table}

\begin{figure}[!hbt]
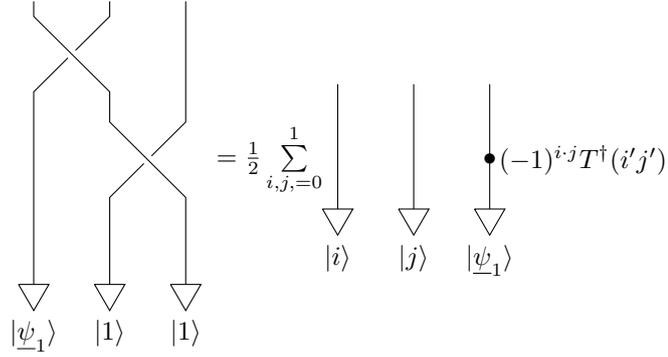

	\begin{equation*}
		\tikz[baseline]{
			%	\draw(0.9,2)rectangle(2.1,0.8);
			%	\draw(1.9,0.6)rectangle(3.1,-0.6);
			\draw(1,2.1)--(1,1.9)--(2,0.9)--(2,0.5)--(3,-0.5)--(3,-1.6536);
			\draw(2.8,-1.6536)--(3.2,-1.6536)--(3,-2)--(2.8,-1.6536);
			\draw(2,2.1)--(2,1.9)--(1.55,1.45);
			\draw(1.45,1.35)--(1,0.9)--(1,-1.6536);
			\draw(0.8,-1.6536)--(1.2,-1.6536)--(1,-2)--(0.8,-1.6536);
			\draw(3,2.1)--(3,0.5)--(2.55,0.05);
			\draw(2.45,-0.05)--(2,-0.5)--(2,-1.6536);
			\draw(1.8,-1.6536)--(2.2,-1.6536)--(2,-2)--(1.8,-1.6536);
			\draw(4.1,0)coordinate(A) node {$=\frac{1}{2}\sum\limits_{i,j=0}^1$};
			\draw(5,1)--(5,-0.6536);
			\draw(4.8,-0.6536)--(5.2,-0.6536)--(5,-1)--(4.8,-0.6536);
			\draw(6,1)--(6,-0.6536);
			\draw(5.8,-0.6536)--(6.2,-0.6536)--(6,-1)--(5.8,-0.6536);
			\draw(7,1)--(7,-0.6536);
			\draw(6.8,-0.6536)--(7.2,-0.6536)--(7,-1)--(6.8,-0.6536);
			\draw(1,-2)coordinate(A) node[below] {$|\underline{\psi}_1\rangle$};
			\draw(2,-2)coordinate(A) node[below] {$|1\rangle$};
			\draw(3,-2)coordinate(A) node[below] {$|1\rangle$};
			\draw(5,-1)coordinate(A) node[below] {$|i\rangle$};
			\draw(6,-1)coordinate(A) node[below] {$|j\rangle$};
			\draw(7,-1)coordinate(A) node[below] {$|\underline{\psi}_1\rangle$};
			\draw(7,0)coordinate(A) node {$\bullet$};
			\draw(7,0)coordinate(A) node[right] {$(-1)^{i\cdot j} T^\dagger(i^\prime j^\prime) $};
		}
	\end{equation*}
	\caption{Diagrammatic representation of the braid teleportation equation 
		\eqref{diagrammatic braid teleportation single qubit}.}
	\label{diagrammatic braid teleportation single qubit22}
\end{figure}

To recover the teleportation equation \eqref{teleportation equation 2}, set \(k^\prime=m^\prime=0\); 
then \(k=m=\frac {1+\eta^l} 2\). Taking \(\epsilon^l=-1\) and \(\eta^l=1\), we have \(k=m=1\). 
For the case \(\epsilon^r=-1\) and \(\eta^r=1\), the braid teleportation equation 
\eqref{braid teleportation single qubit} takes the form 
\begin{equation}
	(B(1,-1)\otimes 1\!\!1_2)(1\!\!1_2\otimes B(-1,1))
	(  |\underline{\psi}_1\rangle \otimes|11\rangle)=\frac{1}{2}\sum_{i,j=0}^1|ij\rangle\otimes 
	U^{-1 1, -1 1}_{00,i^\prime j^\prime}	|\underline{\psi}_1\rangle,
\end{equation}
where the unitary correction operators \(U^{-1 1, -1 1}_{00,ij}\) are given by 
\begin{equation}
	U^{-1 1, -1 1}_{00,i^\prime j^\prime}=(-1)^{i\cdot (j\oplus 1)}    T^\dagger(i^\prime j^\prime) 
\end{equation}
with  \(i^\prime=j\oplus 1\) and \(j^\prime=i \oplus j\). In the other case, \(\epsilon^r=1\) and \(\eta^r=-1\), the braid teleportation equation \eqref{braid teleportation single qubit} becomes 
\begin{equation}
	\label{diagrammatic braid teleportation single qubit}
	(B(-1,1)\otimes 1\!\!1_2)(1\!\!1_2\otimes B(-1,1))
	(  |\underline{\psi}_1\rangle \otimes|11\rangle)=\frac{1}{2}\sum_{i,j=0}^1|ij\rangle\otimes 
	(-1)^{i\cdot j}    T^\dagger(i^\prime j^\prime) 	|\underline{\psi}_1\rangle,
\end{equation}
with  \(i^\prime=j\) and \(j^\prime=i \oplus j\). A diagrammatic representation of the above braid 
teleportation equation is depicted in Fig.~\ref{diagrammatic braid teleportation single qubit22}, 
where the pictorial Yang--Baxter gate \(B(-1, 1)\) from Fig.~\ref{diagrammatic braiding gates} is used twice. 

Note that the connection between the extended Temperley--Lieb diagrammatic approach and the braid teleportation approach to quantum teleportation of a single qubit has been clearly demonstrated in the literature; see, for example, \cite{Zhang2006,Zhang2009, ZZP2015, ZZ2017}.

\subsection{Braid teleportation of multiple qubits}

We now generalize the braid teleportation equation \eqref{braid teleportation single qubit} to the transmission of multiple qubits. First, we reformulate the teleportation equation \eqref{n-qubit teleportation equation11} with \(M=1\!\!1_2^{\otimes n}\):
\begin{equation}
	|\underline{\psi}_n\rangle \otimes |\mathcal{B}_{2n}(\underline{a^\prime b^\prime}) \rangle
	=\frac {1} {2^n} \sum_{\underline{\alpha}^\prime,\underline{\beta}^\prime}
	|\mathcal{B}_{2n}(\underline{\alpha^\prime\beta^\prime})\rangle \otimes 	T_n^\dagger(\underline{a^\prime b^\prime})	
	T_n^\dagger(\underline{\alpha^\prime\beta^\prime}) |\underline{\psi}_n\rangle.
\end{equation} 
Second, the \(n\)-bit strings \(\underline{a}^\prime\), \(\underline{b}^\prime\), \(\underline{\alpha}^\prime\), and \(\underline{\beta}^\prime\) are respectively obtained by bijective mappings between the bits \(a_i^\prime, b_i^\prime\) and \(a_i, b_i\):
\begin{equation}
	a^\prime_i=a_i \oplus \frac{|\epsilon^l_i-\eta^l_i|}{2} b_i^\prime \oplus  \frac{1+\eta^l_i}{2}, \quad  
	b^\prime_i=a_i \oplus b_i 
\end{equation}
with \(\epsilon^l_i=\eta^l_i=\pm 1\), and between the bits \(\alpha_i^\prime, \beta_i^\prime\) and \(\alpha_i, \beta_i\):
\begin{equation}
	\alpha^\prime_i=\alpha_i \oplus \frac{|\epsilon^r_i-\eta^r_i|}{2} \beta_i^\prime \oplus  \frac{1+\eta^r_i}{2}, \quad  
	\beta^\prime_i=\alpha_i \oplus \beta_i 
\end{equation}
with \(\epsilon^r_i=\eta^r_i=\pm 1\). Third, we introduce a family of multiple qubit Yang--Baxter gates, constructed by twisting \(n\)-fold tensor products of the two-qubit Yang--Baxter gates:
\begin{equation}
	\mathcal{B}_{2n}(\underline{\epsilon}^l,\underline{\eta}^l)=\tau_{2n}\bigotimes_{i=1}^n 
	B(\epsilon_i^l,\eta_i^l),\quad 	\mathcal{B}_{2n}(\underline{\epsilon}^r,\underline{\eta}^r)=\tau_{2n}\bigotimes_{i=1}^n 
	B(\epsilon_i^r,\eta_i^r),
\end{equation}
with \(n\)-bit strings \(\underline{\epsilon}^l,\underline{\eta}^l\) and \(\underline{\epsilon}^r,\underline{\eta}^r\). Given the product basis states 
\begin{equation}
	|\underline{a b}\rangle
	=\bigotimes_{i=1}^n |a_i b_i\rangle,\quad  
	|\underline{\alpha\beta}\rangle
	=\bigotimes_{i=1}^n |\alpha_i\beta_i\rangle,
\end{equation} 
the corresponding braid teleportation equation is then formulated as
\begin{equation}
	\label{braid teleportation equation n qubits}
	(\mathcal{B}^\dagger_{2n}(\underline{\epsilon}^r,\underline{\eta}^r)\otimes 1\!\! 1_2^{\otimes n} )
	( 1\!\! 1_2^{\otimes n} \otimes \mathcal{B}_{2n}(\underline{\epsilon}^l,\underline{\eta}^l))	(|\underline{\psi}_n\rangle \otimes  | \underline{a b}\rangle) 
	=\frac {1} {2^n} \sum_{\underline{\alpha},\underline{\beta}}
	|\underline{\alpha\beta}\rangle \otimes 
	U^{\underline{\epsilon}^l \underline{\eta}^l,\underline{\epsilon}^r \underline{\eta}^r}_{
		\underline{a}^\prime \underline{b}^\prime,\underline{\alpha}^\prime \underline{\beta}^\prime}
	|\underline{\psi}_n\rangle,
\end{equation} 
where the unitary correction operators 
\( U^{\underline{\epsilon}^l \underline{\eta}^l,\underline{\epsilon}^r \underline{\eta}^r}_{
	\underline{a}^\prime \underline{b}^\prime,\underline{\alpha}^\prime \underline{\beta}^\prime}\)
are given by 
\begin{equation}
	U^{\underline{\epsilon}^l \underline{\eta}^l,\underline{\epsilon}^r \underline{\eta}^r}_{
		\underline{a}^\prime \underline{b}^\prime,\underline{\alpha}^\prime \underline{\beta}^\prime} 
	= (-1)^{\sum_{i=1}^n f(\epsilon_i^l,\eta_i^l,a_i, b_i)} 
	(-1)^{ \sum_{i=1}^n f(\epsilon_i^r,\eta_i^r,\alpha_i, \beta_i)} 
	T_n^\dagger(\underline{a^\prime b^\prime})	
	T_n^\dagger(\underline{\alpha^\prime\beta^\prime}).
\end{equation}

Consider the case \(a^\prime_i=b^\prime_i=0\), i.e., \(a_i=b_i=\frac {1+\eta^l_i}{2}\). Taking \(\epsilon_i^l=-1\) and \(\eta_i^l=1\) gives \(a_i=b_i=1\). When \(\epsilon^r_i=-1\) and \(\eta^r_i=1\), the braid teleportation equation \eqref{braid teleportation equation n qubits} becomes 
\begin{equation}
	\begin{aligned}
		&( B^{\otimes n}(1,-1) \tau_{2n}^\dagger  \otimes 1\!\! 1_2^{\otimes n} ) 
		( 1\!\! 1_2^{\otimes n} \otimes \tau_{2n} B^{\otimes n}(-1,1)   )
		(|\underline{\psi}_n\rangle \otimes  |11\rangle^{\otimes n}) \\ 
		&=\frac {1} {2^n} \sum_{\underline{\alpha},\underline{\beta}}
		|\underline{\alpha\beta}\rangle \otimes 
		(-1)^{ \sum_{i=1}^n \alpha_i\cdot (\beta_i\oplus 1)} 
		T_n^\dagger(\underline{\alpha^\prime\beta^\prime})
		|\underline{\psi}_n\rangle,
	\end{aligned}
\end{equation} 
with \(\alpha_i^\prime=\beta_i \oplus 1\) and \(\beta^\prime_i=\alpha_i \oplus \beta_i\). 
When \(\epsilon^r_i=1\) and \(\eta^r_i=-1\), the braid teleportation equation \eqref{braid teleportation equation n qubits} takes the form
\begin{equation}
	\label{braid teleportation equation n qubits22}
	\begin{aligned}
		&( B^{\otimes n}(-1,1) \tau_{2n}^\dagger  \otimes 1\!\! 1_2^{\otimes n} ) 
		( 1\!\! 1_2^{\otimes n} \otimes \tau_{2n} B^{\otimes n}(-1,1)   )
		(|\underline{\psi}_n\rangle \otimes  |11\rangle^{\otimes n}) \\ 
		&=\frac {1} {2^n} \sum_{\underline{\alpha},\underline{\beta}}
		|\underline{\alpha\beta}\rangle \otimes 
		(-1)^{ \sum_{i=1}^n \alpha_i\cdot \beta_i} 
		T_n^\dagger(\underline{\alpha^\prime\beta^\prime})
		|\underline{\psi}_n\rangle,
	\end{aligned}
\end{equation} 
with \(\alpha_i^\prime=\beta_i\) and \(\beta^\prime_i=\alpha_i \oplus \beta_i\). For \(n=2\), the above braid 
teleportation equation is illustrated in Fig.~\ref{diagrammatic braid teleportation n qubit22}, 
where the diagrammatic Yang--Baxter gate \(B(-1, 1)\) from Fig.~\ref{diagrammatic braiding gates}, the SWAP gate from Fig.~\ref{fig: graphic twist operator four}, and a graphical two-qubit state \(|\underline{\psi}_2\rangle\) depicted as a product state of single-qubit states from Fig.~\ref{fig: graphic multi qubit product state} are used.  
\begin{figure}[!hbt]
	\begin{equation*}
		%	\label{BBB}
		\scalebox{0.8}[0.9]
		{
			\tikz[baseline]{
				\draw(1,1.9)--(1,0.5)--(2,-0.5)--(2,-3.6536);
				\draw(2,1.9)--(2,1.7)--(3,0.7)--(3,0.5)--(4,-0.5)--(4,-0.9)--(5,-1.9)--(5,-2.1)--(6,-3.1)--(6,-3.6536);
				\draw(3,1.9)--(3,1.7)--(2,0.7)--(2,0.5)--(1.55,0.05);
				\draw(1.45,-0.05)--(1,-0.5)--(1,-3.6536);
				\draw(4,1.9)--(4,0.5)--(3.55,0.05);
				\draw(3.45,-0.05)--(3,-0.5)--(3,-2.1)--(4,-3.1)--(4,-3.6536);
				\draw(2.5,1.2)circle (0.2);
				\draw(0.8,-3.6536)--(1.2,-3.6536)--(1,-4)--(0.8,-3.6536);
				\draw(1.8,-3.6536)--(2.2,-3.6536)--(2,-4)--(1.8,-3.6536);
				\draw(4.8,-3.6536)--(5.2,-3.6536)--(5,-4)--(4.8,-3.6536);
				\draw(5.8,-3.6536)--(6.2,-3.6536)--(6,-4)--(5.8,-3.6536);
				\draw(5,1.9)--(5,-0.9)--(4,-1.9)--(4,-2.1)--(3.55,-2.55);
				\draw(3.45,-2.65)--(3,-3.1)--(3,-3.6536);
				\draw(2.8,-3.6536)--(3.2,-3.6536)--(3,-4)--(2.8,-3.6536);
				\draw(6,1.9)--(6,-2.1)--(5.55,-2.55);
				\draw(5.45,-2.65)--(5,-3.1)--(5,-3.6536);
				\draw(3.8,-3.6536)--(4.2,-3.6536)--(4,-4)--(3.8,-3.6536);
				\draw(4.5,-1.4)circle (0.2);
				\draw(7,-1.1)coordinate(A) node {$=\frac{1}{4}\sum\limits_{\alpha_i,\beta_i=0}^1$};
				\draw(8,-0.1)--(8,-1.7536);
				\draw(7.8,-1.7536)--(8.2,-1.7536)--(8,-2.1)--(7.8,-1.7536);
				\draw(9,-0.1)--(9,-1.7536);
				\draw(8.8,-1.7536)--(9.2,-1.7536)--(9,-2.1)--(8.8,-1.7536);
				\draw(10,-0.1)--(10,-1.7536);
				\draw(9.8,-1.7536)--(10.2,-1.7536)--(10,-2.1)--(9.8,-1.7536);
				\draw(11,-0.1)--(11,-1.7536);
				\draw(10.8,-1.7536)--(11.2,-1.7536)--(11,-2.1)--(10.8,-1.7536);
				\draw(12,-0.1)--(12,-1.7536);
				\draw(11.8,-1.7536)--(12.2,-1.7536)--(12,-2.1)--(11.8,-1.7536);
				\draw(13,-0.1)--(13,-1.7536);
				\draw(12.8,-1.7536)--(13.2,-1.7536)--(13,-2.1)--(12.8,-1.7536);
				\draw(1.5,-4)coordinate(A) node[below] {\(|\underline{\psi}_2\rangle\)};
				\draw(3,-4)coordinate(A) node[below] {$|1\rangle$};
				\draw(4,-4)coordinate(A) node[below] {$|1\rangle$};
				\draw(5,-4)coordinate(A) node[below] {$|1\rangle$};
				\draw(6,-4)coordinate(A) node[below] {$|1\rangle$};
				\draw(8,-2.1)coordinate(A) node[below] {$|\alpha_1\rangle$};
				\draw(9,-2.1)coordinate(A) node[below] {$|\beta_1\rangle$};
				\draw(10,-2.1)coordinate(A) node[below] {$|\alpha_2\rangle$};
				\draw(11,-2.1)coordinate(A) node[below] {$|\beta_2\rangle$};
				\draw(12.5,-2.1)coordinate(A) node[below] {\(|\underline{\psi}_2\rangle\)};
				\draw(12,-0.4)coordinate(A) node {$\bullet$};
				\draw(12,0.1)coordinate(A) node[right] {\((-1)^{ \alpha_1\cdot \beta_1} T^\dagger(\alpha^\prime_1\beta^\prime_1)\)};
				\draw(13,-1.1)coordinate(A) node {$\bullet$};
				\draw(13.2,-1.1)coordinate(A) node[right] {\((-1)^{ \alpha_2\cdot \beta_2} T^\dagger(\alpha^\prime_2\beta^\prime_2)\)};
				%	\draw(12,-1.1)--(13,-1.1);
			}
		}
	\end{equation*}
	\caption{Diagrammatic representation of the braid teleportation equation 
		\eqref{braid teleportation equation n qubits22}.}
	\label{diagrammatic braid teleportation n qubit22}
\end{figure}

In addition to the twisted Yang--Baxter gates \(\mathcal{B}_{2n}(\underline{\epsilon}^l,\underline{\eta}^l)\)
and \(\mathcal{B}_{2n}(\underline{\epsilon}^r,\underline{\eta}^r)\), another family of twisted Yang--Baxter gates can be constructed as
\begin{equation}
	\label{the multiple qubit Yang--Baxter gates}
	\overline{\mathcal{B}}_{2n}(\underline{\epsilon}^l,\underline{\eta}^l)
	=\tau_{2n}\bigotimes_{i=1}^n B(\epsilon_i^l,\eta_i^l)\,\tau^\dagger_{2n},\quad 	\overline{\mathcal{B}}_{2n}(\underline{\epsilon}^r,\underline{\eta}^r)
	=\tau_{2n}\bigotimes_{i=1}^n B(\epsilon_i^r,\eta_i^r)\, \tau^\dagger_{2n}.
\end{equation}
The essential difference between the two types of twisted Yang--Baxter gates is that \(\overline{\mathcal{B}}_{2n}\) is a solution of the Yang--Baxter equation, whereas \(\mathcal{B}_{2n}\) is not. In terms of \(\overline{\mathcal{B}}_{2n}\), the braid teleportation equation \eqref{braid teleportation equation n qubits} for the transmission of \(n\) qubits is reformulated as 
\begin{equation}
	(\overline{\mathcal{B}}^\dagger_{2n}(\underline{\epsilon}^r,\underline{\eta}^r)\otimes 1\!\! 1_2^{\otimes n} )
	( 1\!\! 1_2^{\otimes n} \otimes \overline{\mathcal{B}}_{2n}(\underline{\epsilon}^l,\underline{\eta}^l))	(|\underline{\psi}_n\rangle \otimes  | \underline{a}\, \underline{b}\rangle) 
	=\frac {1} {2^n} \sum_{\underline{\alpha},\underline{\beta}}
	|\underline{\alpha}\underline{\beta}\rangle \otimes 
	U^{\underline{\epsilon}^l \underline{\eta}^l,\underline{\epsilon}^r \underline{\eta}^r}_{
		\underline{a}^\prime \underline{b}^\prime,\underline{\alpha}^\prime \underline{\beta}^\prime}
	|\underline{\psi}_n\rangle,
\end{equation} 
with \(| \underline{a}\, \underline{b}\rangle =\tau_{2n} | \underline{a b}\rangle\) and \(|\underline{\alpha}\underline{\beta}\rangle =\tau_{2n} |\underline{\alpha\beta}\rangle\).

\section{Concluding remarks}

This paper has investigated the algebraic properties of generalized Bell states and provided a topological description of quantum teleportation. Motivated by the question of when the extension of a generalized Bell basis by a matrix preserves orthonormality, we conducted a detailed study of the basis group and basis theorem. A generalized multi‑qubit Bell state is obtained by applying the twist operator to a tensor product of two‑qubit Bell states. A topological description of quantum teleportation for multiple qubits is given via the extended Temperley–Lieb diagrammatic approach, together with diagrammatic SWAP gates. Furthermore, braid teleportation for multiple qubits is described using Yang–Baxter gates and SWAP gates.

In addition to the above, we have carefully examined observables, quantum circuits, and entanglement measures associated with generalized Bell states. The paper thus presents a detailed study of Bell states and quantum teleportation, which is valuable for research in quantum information and computation. Both the extended Temperley–Lieb diagrammatic approach and the braid group approach are expected to play important roles in quantum communication \cite{BCZ1998}, quantum networks \cite{Kimble2008}, and measurement based quantum computation \cite{ZZP2015, ZZ2017}. Furthermore, the investigation of Bell states and quantum teleportation via Yang–Baxter gates sheds light on integrable quantum computation \cite{Zhang2013, ZW2022}, defined as universal quantum computation achieved through Yang–Baxter gates. Finally, the topological descriptions of Bell states and quantum teleportation clearly reveal a fundamental structure governing quantum entanglement and measurements, which is crucial for understanding the counterintuitive nature of quantum mechanics \cite{EPR35, Bohm51, Bell64}.

Future work on various aspects of Bell states and quantum teleportation will be submitted elsewhere. First, since generalized Pauli matrices \eqref{general pauli x} and \eqref{general pauli z} are considerably more intricate than the standard Pauli matrices \eqref{pauli x z}, it is necessary to examine the basis transformation between generalized two‑qudit Bell states (constructed via generalized Pauli matrices) and generalized multiple qubit Bell states (constructed via tensor products of Pauli matrices). Second, it would be interesting to perform a thorough study of the concurrence formula \eqref{concurrence formula} for multiple qubit entangled states in the context of quantum information and computation, as well as to explore the coherent state representation \cite{Fujii2001,Fivel2002,Fujii2001a} of generalized Bell states. Third, it is meaningful to construct a suitable illustration for a general multiple qubit entangled pure state (instead of Fig.~\ref{fig: graphic multi qubit entangled state}) and to provide an appropriate diagrammatic representation of quantum teleportation using non‑maximally entangled states \cite{MH1999}. Fourth, with reference to Fig.~\ref{diagrammatic braid teleportation n qubit22}, an investigation integrating the extended Brauer diagrammatic approach \cite{Brauer1937} and multiple qubit Yang–Baxter gates \eqref{the multiple qubit Yang--Baxter gates} is expected to yield a more compelling topological description of quantum teleportation using generalized multiple qubit Bell states. Crucially, in applying our algebraic and topological approach, we seek novel protocols and useful classes of entangled states for quantum information and computation and look forward to further developments.

\section*{Acknowledgements}

Yong Zhang was supported by the NSF of China (Grant No. 11574237) for his work on Integrable Quantum Computation \cite{Zhang2013, ZW2022}. He is greatly  indebted to Professor Mo-Lin Ge and Professor Louis H. Kauffman  for a first study on  the application of the quantum Yang--Baxter equation to quantum computation \cite{ZKG04, ZKG05}. He also thanks the participants in the courses on Quantum Field Theory and Quantum Information \& Computation   in Wuhan University for their helpful discussions during the period from 2012 to 2025. He also sincerely thanks  anonymous reviewers for their reports, which have helped to improve the quality of this manuscript.

\section*{Author Contributions Statement}

Ming Lian conducted an initial study of generalized \(2n\)-qubit Bell states and quantum teleportation in his Bachelor's thesis (supervised by Yong Zhang). Wei Zeng further investigated these subjects in his Master's thesis (also supervised by Yong Zhang) and acknowledges Yan-Zhen Hao for valuable assistance with the Matlab code. Yong Zhang then performed a comprehensive reinvestigation of the generalized Bell bases and quantum teleportation, and wrote the present manuscript.

\section*{Data Availability Statement}
   
 Data sharing was not applicable to this article as no datasets were generated or analyzed during the current study.

\appendix

\section{Proof of Lemma \ref{lemmainductiveproof} by induction}
\label{lianmingproof}

Regarding the proof of Lemma~\ref{lemmainductiveproof}, the simplest approach is a direct application of the orthonormal basis $T_n(\underline{\alpha\beta})$ with respect to the Hilbert--Schmidt inner product, as previously described in Subsection~\ref{basis-theorem-multi-qubit}. To exhibit the elegant properties of tensor products of Pauli gates, we present an algebraic proof by induction in the following. Rather than proving the original lemma directly, we first propose another lemma.  
\begin{lemma}	\label{lemmainductiveproofdeform}
	The  constraint condition on $T_n(\underline{\alpha\beta})$ given by
	\begin{equation}
		\frac {1} {2^n} \operatorname{tr}\bigl(I(n) \, T_n(\underline{\alpha\beta}) \bigr)
		=\delta_{\underline{\alpha}\,\underline{0}}
		\delta_{\underline{\beta}\,\underline{0}},
	\end{equation}
	is satisfied if and only if  $I(n)=1\!\! 1_{2^n}$, while the condition 
\begin{equation}
	\frac {1} {2^n} \operatorname{tr}\bigl(\tilde{I}(n) \, T_n(\underline{\alpha\beta}) \bigr)
	=0,
\end{equation}
	is satisfied if and only if  $\tilde{I}(n)=0$.
\end{lemma}

At $n=1$, the constraints on the $2\times 2$ matrix elements $I_{ij}(1)$, $i, j=0, 1$, are
\begin{equation}
	\begin{aligned}
		&\operatorname{tr}(I(1))=I_{00}(1)+I_{11}(1)=2,\\
		&\operatorname{tr}(I(1)X)=I_{01}(1)+I_{10}(1)=0,\\
		&\operatorname{tr}(I(1)ZX)=-I_{01}(1)+I_{10}(1)=0,\\
		&\operatorname{tr}(I(1)Z)=I_{00}(1)-I_{11}(1)=0,
	\end{aligned}
\end{equation}
so that $I(1)=1\!\!1_2$, and the $4\times 4$ matrix equations for the elements $\tilde{I}_{ij}(1)$ take the form
	\begin{equation}
		\begin{pmatrix}1&0&0&1\\0&1&1&0\\0&-1&1&0\\1&0&0&-1\end{pmatrix}
		\begin{pmatrix} \tilde{I}_{00}(1)\\ \tilde{I}_{01}(1)\\ 
			\tilde{I}_{10}(1) \\ \tilde{I}_{11}(1)  \end{pmatrix}
		=\begin{pmatrix}0\\0\\0\\0\end{pmatrix},
	\end{equation}
	so that $\tilde{I}(1)=0$.
	
	Suppose the lemma holds in the $n$-qubit Hilbert space. We now prove it in the $2^{n+1}$ dimensional Hilbert space. Write $I(n+1)$ as a $4\times 4$ block matrix,
	$$I(n+1)=\begin{pmatrix}  I_{00}(n)&  I_{01}(n)\\  I_{10}(n)&  I_{11}(n)  \end{pmatrix}$$
	with $2^n\times 2^n$ matrices $I_{ij}(n)$, $i, j=0, 1$.  
	 
	  When both $\underline{\alpha}$ and $\underline{\beta}$ are zero $n$-bit strings, the constraint equations expand to 
	 \begin{equation}
	 	\label{1}\begin{aligned}
	 		&\operatorname{tr}\bigl(I(n+1)(1\!\!1_2\otimes  T_n(\underline{00}) )\bigr)=2^{n+1},\\
	 		&\operatorname{tr}\bigl(I(n+1)(X\otimes T_n(\underline{00}))\bigr)=0,\\
	 		&\operatorname{tr}\bigl(I(n+1)(ZX\otimes T_n(\underline{00}))\bigr)=0,\\
	 		&\operatorname{tr}\bigl(I(n+1)(Z\otimes T_n(\underline{00}))\bigr)=0,
	 	\end{aligned}
	 \end{equation}
 which yield constraints on $I_{00}(n)$ and $I_{11}(n)$,
 \begin{equation}
 	\operatorname{tr}\bigl(I_{00}(n)  T_n(\underline{00}) \bigr)=\operatorname{tr}\bigl(I_{11}(n)  T_n(\underline{00}) \bigr)=2^n, 
 \end{equation}
 and on  $I_{01}(n)$ and $I_{10}(n)$, 
 \begin{equation}
 	\operatorname{tr}\bigl(I_{01}(n)  T_n(\underline{00}) \bigr)=\operatorname{tr}\bigl(I_{10}(n)     T_n(\underline{00}) \bigr)=0.
 \end{equation}
	
 When either $\underline{\alpha}$ or $\underline{\beta}$ is a nonzero $n$-bit string, the constraint 
 equations are 
 \begin{equation}
 \label{2}\begin{aligned}
 	&\operatorname{tr}\bigl(I_{00}(n)T_n(\underline{\alpha\beta})\bigr)+\operatorname{tr}\bigl(I_{11}(n) T_n(\underline{\alpha\beta})\bigr)=0,\\
 	&\operatorname{tr}\bigl(I_{01}(n) T_n(\underline{\alpha\beta})\bigr)+\operatorname{tr}\bigl(I_{10}(n) T_n(\underline{\alpha\beta})\bigr)=0,\\
 	&-\operatorname{tr}\bigl(I_{01}(n) T_n(\underline{\alpha\beta})\bigr)+\operatorname{tr}\bigl(I_{10}(n) T_n(\underline{\alpha\beta})\bigr)=0,\\
 	&\operatorname{tr}\bigl(I_{00}(n) T_n(\underline{\alpha\beta})\bigr)-\operatorname{tr}\bigl(I_{11}(n) T_n(\underline{\alpha\beta})\bigr)=0,
 	\end{aligned}
 \end{equation} 
 which yield constraints on $I_{00}(n)$ and $I_{11}(n)$,
 \begin{equation}
 	\operatorname{tr}\bigl( I_{00}(n) T_n(\underline{\alpha\beta})\bigr)=\operatorname{tr}\bigl( I_{11}(n)  T_n(\underline{\alpha\beta})\bigr)=0,
 \end{equation}
 and on  $I_{01}(n)$ and $I_{10}(n)$, 
 \begin{equation}
 	\operatorname{tr}\bigl(I_{01}(n) T_n(\underline{\alpha\beta})\bigr)=\operatorname{tr}\bigl(I_{10}(n) T_n(\underline{\alpha\beta})\bigr)=0.
 \end{equation}
 
Therefore, by the induction hypothesis, one obtains 
$I_{00}(n)=I_{11}(n)=1\!\!1_2^{\otimes n}$ and $I_{01}(n)=I_{10}(n)=0$, so that 
$I(n+1)=1\!\!1_2^{\otimes (n+1)}$. Similarly, one can verify that $\tilde{I}(n+1)=0$. Consequently, Lemma~\ref{lemmainductiveproof} is established.	
	
%%%%%%%%%%%%%%%%%%%%%%
%%%%%%%%%%%%%%%%%%%%
%%%%%%%%%%%%%%%%%%%

%%%%%%%%%%%%%%
%%%%%%%%%%%%%

\bibliographystyle{unsrt} %参考文献样式, plain,unsrt,alpha,abbrv,chinesebst

\end{document}